\documentclass[%
 reprint,
 superscriptaddress,
 amsmath,amssymb,
 aps,
]{revtex4-2}

\usepackage{graphicx}
\usepackage{dcolumn}
\usepackage{bm}
\usepackage[utf8]{inputenc} 
\usepackage{color}
\usepackage{etoolbox} 
\usepackage{hyperref}
\usepackage{natbib} 
\usepackage{ulem} 
\usepackage[english]{babel}

\newcommand{\oncepermin}{1/min.}

\begin{document}

\preprint{APS/123-QED}

\title{Development Considerations for High-Repetition-Rate HEDP Experiments}

\author{Scott Feister}
\affiliation{Department of Physics \& Astronomy, University of California Los Angeles, Los Angeles, USA}
\affiliation{Department of Computer Science, California State University Channel Islands, Camarillo, USA}

\author{Patrick L. Poole}
\affiliation{Lawrence Livermore National Laboratory, Livermore, USA}

\author{Peter V. Heuer}
\affiliation{Department of Physics \& Astronomy, University of California Los Angeles, Los Angeles, USA}

\date{\today}

\begin{abstract}
This paper discusses experimental techniques and considerations associated with the transition to high repetition-rate experiments in High Energy Density Physics (HEDP). We particularly highlight approaches to experimentation that become practical only at a threshold of repetition rate. We focus on the transition from operation at several-shots-per-day towards operation in the range of \oncepermin{} to 1~Hz.
\end{abstract}

\maketitle

\section{Introduction}

A new generation of high-repetition-rate (HRR) lasers are being proposed and developed for use in the field of High Energy Density Physics (HEDP) and beyond \cite{brocklesby_ican_2014, Roso2018high, Bayramian2007mercury, Schaeffer2018platform}. 

While the highest energy facilities may always necessarily operate at low shot rates, these facilities can now be complemented by high repetition rate lasers that achieve similar intensities at lower energies and shorter pulse lengths.

The ability to conduct experiments at HRR enables new types of experimental designs to collect large multi-dimensional, high-statistic datasets. These datasets are ideal for the development and application of machine learning~\cite{Spears2018deep} and statistical techniques~\cite{Gopalaswamy2019tripled}. Large sample sizes also enable the stochastic study of complex nonlinear systems (e.g. turbulence) and rare or low-signal events. A successful transition to high-repetition-rate HEDP experiments provides many new experimental modes which can be leveraged to answer many sorts of scientific questions and to prototype real-world applications requiring high-average-power \cite{sinars_role_2015, malka_principles_2008, gales_laser_2015, kitagawa_hi-rep._2013, masood2016}. Conducting these types of experiments will require the use of next-generation targetry, detector technology and data acquisition systems. In addition, new data analysis techniques, and experimental logistics must be developed and adopted. 


This paper comprises a high-level discussion of the experimental techniques and challenges associated with the transition from low repetition rate (LRR, several shots/day)  HEDP experiments towards experiments at high repetition rate (HRR, \oncepermin{} to 1~Hz). The authors each have several years of experience with successful planning and execution of HEDP experiments in this repetition-rate range.

While the information in this paper is tailored to HEDP experiments with repetition rates between \oncepermin{} and 1~Hz,  much of it is applicable to much higher repetition rates. However, additional problems that may arise in HEDP experiments faster than $\sim$1~Hz, observed by the authors and also in the literature in $\sim$1~kHz systems (e.g. environmental interference between shots, synchronization of single-shot data, network-limited data pipelines) are outside the scope of this paper. We also will not attempt to motivate specific science experiments or applications, nor to give a survey of the areas of physics research which will be enhanced by the next generation of high-repetition-rate lasers. See for example the National Academy of Science report on bright light \cite{NAS_2019} or some sections of the Extreme Light Infrastructure whitebook \cite{ELI_whitebook}.

Our aim in this paper is to stimulate HRR work in the area of HEDP by providing an overview of the possible design space and technology, both currently available and under development. We will discuss relevant considerations for HRR HEDP experiments in the areas of targets, detectors, experimental approaches, and data analysis.

\section{Targets}
For a LRR system, targets are typically designed and built on a one-off basis. For $\leq$\oncepermin{} systems, simple targets (e.g. foils, solids) may be prepared ahead of time, with a fresh surface being presented before each shot. However, more complicated targets must be fabricated at repetition rate. The target fabrication itself must fit in or near the chamber, be fast, and be automated. Researchers have developed and are actively exploring approaches suitable for high-repetition-rate operation. An excellent review of the state-of-the-art in high-repetition-rate target development for HEDP is found in \citet{Prencipe2017targets}.

A high repetition rate cannot be maintained if the target mechanism is destroyed after a shot. As energies increase, it will become necessary to design target mechanisms that either operate at sufficient distance from or can survive the explosions typical of HEDP experiments. This distance can depend strongly on laser energy and intensity and target constituents, usually on the few mm scale for conditions mostly likely to be seen in near-future HRR facilities ($<$ 100 J, $>$ 10$^{20}$ W/cm$^2$) \cite{grim_foam_2018, poole_LCdamage_2019}.

Finding ways to create complex target geometries and structures ``on-the-fly'' will also be important for future experiments, and innovations in this area is a major opportunity for growth. As with targets at lower rep rates, long pulse lasers can be used to shape and "assemble" the target further in the moments before the main experiment \cite{lubcke_prospects_2017}. This and other techniques (e.g. applying external magnetic fields, adding high-density nanoparticles to a low-density fluid, structuring gas target nozzles \cite{grittani_high_2014}, or colliding multiple high-speed liquids to create interesting geometries \cite{hoath_atomization_2010}) can help add complexity to high-repetition-rate targets.

While the development of HRR target formation, alignment, and protection from fratricide (target quality reduction or full destruction by laser interaction with its neighbor) is critical to next generation laser applications, it also progresses the related area of intensity contrast enhancement via plasma mirror interaction. Here the laser is focused onto a sacrificial optic region such that unwanted pre-pulses transmit while the main pulse is of sufficient fluence to create a plasma that reflects the remainder of the light. Energy is lost in this process, and plasma mirror development has progressed at the single-shot level by improving their efficiency (a figure of merit is 1-2 orders of magnitude reduction in pre-pulse level at the cost of between 50-90\% energy reflection) and the quality of the reflected beam. HRR plasma mirrors present similar challenges to targets but have similar solutions, and some demonstrations exist that could reasonably scale to the 1~Hz scale \cite{poole_plasma_2016, shaw_tape_2016}. Critically, HRR plasma implementation could serve dual purposes by improving laser contrast while also redirecting the preponderance of target debris away from expensive target chamber optics.

Types of target systems with demonstrated application at high-repetition-rate laser facilities include flowing gases \cite{salehi_mev_2017, Aniculaesei2018novel, Lorenz2019characterization}, 
flowing liquids~\cite{george_high_2019, Schwind2019high, stan_liquid_2016, Kim2018development, thoss2003}, microfluidic target assembly~\cite{Inoue2016novel},  static liquid films \cite{poole_liquid_2014, poole_moderate_2016}, cryogenic liquids~\cite{Gauthier2017high} and solids~\cite{Kraft2018first}, pre-assembled-target droppers \cite{komeda_target_2013}, and motorized/rastered targets (e.g. tape-drive targets \cite{noaman-ul-haq_statistical_2017}, spinning disks \cite{hou_mev_2009}, and rotating cylinders~\cite{Schaeffer2018platform}.)

At HRR the contribution of target material from previous shots to the vacuum environment can be non-negligible. Depending on the repetition rate, it may still be possible to remove most of this material in between shots with a sufficiently powerful vacuum system. However, experiments at very high repetition rates ($>$ 1 Hz) will likely be performed in a steady-state equilibrium environment that includes gaseous target material. Ensuring that shots are reproducible under these conditions will require careful monitoring and managing of the gas partial pressures in the chamber throughout the experiment. Shots prior to the establishment of equilibrium, either at the beginning of a run or immediately after a change of targets, should be discarded.

\section{Detectors}
Currently, many HEDP diagnostics are designed under the assumption that the laser repetition-rate is the limiting factor on the repetition-rate of an experiment. As such, analog detectors have been utilized due to their high-dynamic range, compactness, and sensitivity, without adversely affecting the experimental data acquisition rate. 

Modern alternatives to traditional analog detectors, which cannot operate at HRR, are fast digital detectors. Digital detectors have two extremely useful features that cannot be matched by their analog counterparts: they can be used repeatedly shot-after-shot and they afford the possibility of real-time feedback to the experimenter.

Sometimes, appropriate digital detectors are commercially available; for example, in the case of a high-repetition-rate-ready optical spectrometer. However, because HEDP experiments often have exotic detector requirements, many times we must develop our own customized scientific equipment. Detector designs for high-repetition-rate HEDP often include conversion of non-electrical, non-optical signals into electrical or optical signals usable by fast digitizers and digital cameras.

When no suitable alternative to low-repetition-rate detectors can be found, the operational mode of an experiment may still be improved to leverage high repetition-rate. One may think that the experimental repetition-rate must be limited by its slowest detector, but this is not necessarily true. Experimenters may, for example, choose to use digital diagnostics to adjust inputs and find a certain condition, then expose an image plate for several shots under that ``optimal'' condition. This may be a way to bridge into a future where we have created digital implementations of the entire suite of HEDP diagnostics.

An advantage of HRR experiments is that, under certain experimental circumstances, diagnostics can be aligned during full-energy shots with real-time feedback. For example, the positioning of a pinhole could be adjusted to optimize the actual measured signal. Utilizing this capability could allow complex diagnostics to be prepared more quickly.

\subsection{Optical Diagnostics}

Many HEDP diagnostics ultimately rely on an optical detector (eg. camera, diode). Fortunately, digital photon detectors such as fast-gated CCDs, streak cameras, and framing cameras~\cite{Kimbrough2010standard} capable of operating at rates much greater than or equal to 1~Hz are already commonplace in HEDP experiments. These diagnostics would require relatively little modification to transition to HRR.  

\subsection{Particle Diagnostics}
Low repetition rate HEDP experiments commonly use radiochromatic film, image plates, and CR39 plates to record particle fluxes. However, retrieving these plates after shots typically requires venting the target chamber, which is not feasible at high repetition rates. Further, currently existing processing and scanning techniques for these plates have extremely limited throughput, such that scaling them to high repetition rates is untenable. Installing a few of these detectors at a time in the chamber may allow high repetition rate operation for a few seconds or minutes, but would result in a much lower day-averaged shot rate. 

A few examples showcasing the variety of approaches to digital ion detectors are found in \cite{metzkes_online_2016, reinhardt_pixel_2013, prokupek_development_2014, morrison_mev_2018, dover_scintillator_2017, cirrone_transport_2015}. Although the quality and availability of HEDP-relevant digital detectors is increasing, the features that make certain analog detectors popular in HEDP are not easily replicated in a digital fashion. Capturing salient features of radiochromic films, image plates, and CR39 with digital detectors is an area that requires substantial innovation.

\subsection{Other Diagnostics}
Physical probes are typically not an option for HEDP experiments (due to extreme conditions), but may be applicable to non-HEDP experiments performed using an HRR facility. Antennas and electrical probes can directly measure magnetic and electric fields in low-density plasmas with no limitation on repetition rate~\cite{Everson2009design}. 
Another technique occasionally used at LRR is post-shot chemical analysis of either the residual gases in the vacuum chamber or material deposited on a witness plate. This technique would be difficult to operate at HRR, but could be applied to a selective subset of shots.

\section{Experimental Approaches}

High-repetition-rate facilities are capable of experimental approaches entirely outside the range of methods developed and refined for several decades at several-shots-per-day facilities. This section defines and illustrates several categories of experimental approaches well suited to high-repetition-rate HEDP.

\subsection{Parameter Scan}
Input parameters can be changed a large number of times in a single high-repetition-rate experiment. Scans can vary either or both the experimental parameters (e.g. varying the thickness of a target) or the detector settings (e.g. camera timing) to assemble multi-dimensional datasets. We will describe three types of parameter scans that may be found useful in future HEDP experiments, with hypothetical and concrete examples of each.

\subsubsection{Sequenced Parameter Scan}
A pre-sequenced, organized sweep over several values of one (or several) parameters can provide a systematic 1D (or N-D) mapping of a parameter space.

\bigskip \noindent{Examples of a sequenced parameter scan:}
\begin{itemize}
  \setlength\itemsep{0em}
  \item{Incrementally adjusting delay time of either a diagnostic or a secondary backlighter laser to create a high-framerate movie with each frame being a proton radiograph~\cite{Heuer2016fast}.}
  \item{Translating a thin sheet target through best laser-focus, to map out the effect of laser intensity on particle acceleration. (Cf. Fig.~3 of \citet{morrison_mev_2018})}
  \item{Translating the entrance pinhole of a spectrometer in two dimensions to create an energy-resolved image.}
  \item{Repeating time-resolved spectrographic measurements at successive wavelength bands to obtain a time-resolved spectrum.}
\end{itemize}

\subsubsection{Stochastic Parameter Scan}
A stochastic parameter scan explores a wide set of input conditions to statistically sample the parameter space. Unlike a sequenced parameter scan, inputs are not necessarily varied systematically or in a sequence. Sparse random sampling of a wide multidimensional parameter space may be useful in machine learning and/or testing statistical models.

\bigskip \noindent{Examples of a stochastic parameter scan:}
\begin{itemize}
  \setlength\itemsep{0em}
  \item{Embracing large random timing jitter between pump and probe pulses, then sorting results post-facto by delay time to resolve temporal dynamics. (Cf. \citet{wilk_oscat_2011}}.)
  \item{Creating a dataset with laser-A energy, laser-A focal spot size, laser-B energy, laser-B focal spot size, relative timing, pulse shape, pulse duration, target position, electron spectrum, plasma density, and plasma temperature. Computing Pearson coefficients between values in this dataset. (Cf. Fig.~6 of \citet{noaman-ul-haq_statistical_2017} for an example with fewer variables.)}
\end{itemize}

\subsubsection{Optimization Search}
An optimization search has the goal of minimizing, maximizing, or setting outputs to a certain values. This is achieved by adjusting input parameters while observing output values in real time. It might involve, for example, a binary search to narrow in from above and below a target output value.

\bigskip \noindent{Examples of an optimization search:}
\begin{itemize}
  \setlength\itemsep{0em}
  \item{During experimental setup of a pump-probe proton radiography experiment, dynamically adjusting the TNSA-target thickness to achieve a specific proton spectrum.}
  \item{Searching for the critical pressure of a phase transition.}
\end{itemize}

\subsection{Building Statistics}
While the methods in the prior three sub-sections relied on variation of experimental inputs, the methods in the next three sections rely on building statistics by performing many experimental trials with \emph{identical} inputs. We outline and give examples of three distinct and complementary approaches.

\subsubsection{Characterization of Fluctuation}
The characterization of a stochastic plasma physics dynamics becomes possible in high-repetition-rate experiments. One can explore and/or quantify the range of outputs that occur while retaining nominally fixed inputs.

\bigskip \noindent{Examples of characterization of fluctuation:}
\begin{itemize}
  \setlength\itemsep{0em}
  \item{Performing standard deviation analysis of a proton spectrum created by TNSA.}
  \item{Categorizing shadowgraphic images of plasma blowoff that vary due to small fluctuations in laser spatial mode.}
  \item{Studying laser absorption in nanoparticles by recording a single nanoparticle explosion per shot, over many shots with varying features (cf. \citet{hickstein_mapping_2014}, especially Fig.~4})\\
  \item{Performing statistical analysis of turbulent plasma flows.}
\end{itemize}

\subsubsection{Building Signal or Building Counts}
Building signal or building counts is applicable in experiments with low-signal detectors requiring multi-shot averaging, or in experiments leveraging single-hit particle detectors. Similar to the characterization fluctuation, this approach involves repetition with fixed inputs to build up output statistics.

If stochastic plasma physics dynamics are also involved, this method may also benefit from binning the various trials of the experiment into different categories based on a secondary detector (see the last example below).

\bigskip \noindent{Examples of building signal or building counts:}
\begin{itemize}
  \setlength\itemsep{0em}
  \item{Building a high-resolution, energy-resolved 2D-spatial-image through a series of single-hit X-ray camera images.}
  \item{Utilizing a pixel-based single-hit ($\leq$ one particle per pixel per shot) proton detector as described in \citet{reinhardt_pixel_2013}.}
  \item{Building a multi-shot-averaged X-ray spectrum from a subset of shots where the plasma density at Point~A was between 10 and 15.}
\end{itemize}

\subsubsection{Search for Rare Events}
Searches for rare outcomes of an experiment are possible through repetition. The implementation of this approach is similar to when one wishes to characterize of a stochastic process, but with a different goal and with different conditions for ending the experiment. One may note that this scheme may benefit from the addition of real-time-data-analysis so that the success rate of finding rare events is immediately known to the experimenter.

\bigskip \noindent{Example of a search for rare events:}
\begin{itemize}
  \setlength\itemsep{0em}
  \item{In prior data, on just a few shots out of five thousand, a super-high-energy X-ray emission was observed. We hadn't fielded an electron spectrometer, but we hypothesize that the X-ray emission was driven by a process that would have been visible in the electron spectrum. For the next experiment, we field an electron spectrometer and we wait to capture another (or several more) super-high-energy X-ray emission(s).}
\end{itemize}

\subsection{Machine-Learning Feedback Loop}
An advanced application of high-repetition-rate facilities is to yield control of input parameters to machine-learning computer programs. These programs could adjust themselves on-the-fly based on real-time experimental outputs, or they could be pre-trained from prior datasets. The feedback loop is closed in the sense that the outputs feed into the inputs.

\bigskip \noindent{Example of a machine-learning feedback loop:}
\begin{itemize}
  \setlength\itemsep{0em}
  \item{Genetic algorithm used to collimate a laser-produced electron beam via control of a deformable mirror in the laser chain (cf. \citet{he_coherent_2015})}
\end{itemize}

\section{Data Analysis \& Storage}

A single-shot HEDP experiment on a low-repetition rate facility may produce hundreds of gigabytes of raw data~\cite{Hutton2012experiment}, while at even modest repetition rates of \oncepermin, a HRR experiment may collect $>$ 3000 shots in a single experiment day. Even assuming a more modest data rate of 1 GB per shot, such a facility would produce $>$3 terabytes of data per day. In order to provide real-time feedback to inform experimental decisions, this data must be processed by automated analysis routines and made available to experimenters on the same time scale as the repetition rate. Archiving and transporting such large volumes of data may be prohibitive, necessitating some form of automated data reduction. 

It is therefore inescapable that the development of successful HRR HEDP experimental platforms will require the creation of automated analysis software, the adoption of space and memory efficient data formats, and dedicated computing resources (on the scale of a small cluster). These problems have already been solved by other large-scale experimental physics platforms such as the Large Hadron Collider~\cite{Bird2011computing}, and best practices have been established~\cite{Demchenko2012addressing, Katal2013big}.

\subsection{Automated Data Processing for Real-Time Feedback}

Real-time data analysis (e.g. automatically retrieving velocity from a VISAR image, phase reconstruction from an interferometry image) is required to provide feedback during an experiment in order to verify the quality of data being collected and plan for subsequent shots. The volume of data collected at HRR makes in-depth, human-guided analysis of each experimental output and shot impractical. Consequently, it is necessary to develop automated data analysis tools which do not require user input. Large-scale experimental physics platforms, including the National Ignition Facility~\cite{Hutton2012experiment}, have already implemented some level of automated analysis.

In cases where fully automated analysis is not practical, it is worth noting that shot-by-shot analysis can be fully-automated without full automation of hour-by-hour setup. For example, the wavelength parameter of an otherwise automated interferometer image processing program could be changed manually every thousands shots.

\subsection{Data Format}

Data should be stored in a modern hierarchical format such as HDF5~\cite{hdf5}. These data formats offer several features that are useful for handling the large ($>$100 GB) data files that will be produced by HRR experiments and that support chunking, lossless compression, and parallel access. Chunked datasets allow efficient access to parts of a dataset whose entirety exceeds the available system access memory, allowing large datasets to be analyzed on personal computers. Lossless compression algorithms can dramatically reduce file sizes at the cost of extra processing time when reading and writing data. Parallel access allows files residing on a central server to be read simultaneously by multiple users, enabling multiple analysis routines to be run in parallel. 

\section{Environment}
Depending on the experiment, the radiation created by an HRR experiment over the course of a day may require special safety precautions. A discussion of radiation safety considerations in HRR experiments is found in Gizzi2010~\cite{Gizzi2010high}.

\section{Conclusion}
Experiments designed to take advantage of a shot rate of \oncepermin{} to 1~Hz have the opportunity to expand High Energy Density Physics into a computationally-intensive and rich landscape of real-time feedback and ``big data'' statistical analysis. Successful science on novel high-repetition-rate laser facilities will require consideration and development of new targets, detectors, experimental approaches, and data analysis. In this paper, we provided an introduction to experimental techniques and considerations we have found useful in each of those categories. Future work will require drawing on expertise both inside and outside of our field to refine and implement these techniques in scientific experiments, as well as identifying and refining specific scientific goals that may be uniquely achieved with these approaches that are relatively novel within our field.

\bibliography{hrr}

\begin{thebibliography}{55}%
\makeatletter
\providecommand \@ifxundefined [1]{%
 \@ifx{#1\undefined}
}%
\providecommand \@ifnum [1]{%
 \ifnum #1\expandafter \@firstoftwo
 \else \expandafter \@secondoftwo
 \fi
}%
\providecommand \@ifx [1]{%
 \ifx #1\expandafter \@firstoftwo
 \else \expandafter \@secondoftwo
 \fi
}%
\providecommand \natexlab [1]{#1}%
\providecommand \enquote  [1]{``#1''}%
\providecommand \bibnamefont  [1]{#1}%
\providecommand \bibfnamefont [1]{#1}%
\providecommand \citenamefont [1]{#1}%
\providecommand \href@noop [0]{\@secondoftwo}%
\providecommand \href [0]{\begingroup \@sanitize@url \@href}%
\providecommand \@href[1]{\@@startlink{#1}\@@href}%
\providecommand \@@href[1]{\endgroup#1\@@endlink}%
\providecommand \@sanitize@url [0]{\catcode `\\12\catcode `\$12\catcode
  `\&12\catcode `\#12\catcode `\^12\catcode `\_12\catcode `\%12\relax}%
\providecommand \@@startlink[1]{}%
\providecommand \@@endlink[0]{}%
\providecommand \url  [0]{\begingroup\@sanitize@url \@url }%
\providecommand \@url [1]{\endgroup\@href {#1}{\urlprefix }}%
\providecommand \urlprefix  [0]{URL }%
\providecommand \Eprint [0]{\href }%
\providecommand \doibase [0]{https://doi.org/}%
\providecommand \selectlanguage [0]{\@gobble}%
\providecommand \bibinfo  [0]{\@secondoftwo}%
\providecommand \bibfield  [0]{\@secondoftwo}%
\providecommand \translation [1]{[#1]}%
\providecommand \BibitemOpen [0]{}%
\providecommand \bibitemStop [0]{}%
\providecommand \bibitemNoStop [0]{.\EOS\space}%
\providecommand \EOS [0]{\spacefactor3000\relax}%
\providecommand \BibitemShut  [1]{\csname bibitem#1\endcsname}%
\let\auto@bib@innerbib\@empty
\bibitem [{\citenamefont {Brocklesby}\ \emph {et~al.}()\citenamefont
  {Brocklesby}, \citenamefont {Nilsson}, \citenamefont {Schreiber},
  \citenamefont {Limpert}, \citenamefont {Brignon}, \citenamefont
  {Bourderionnet}, \citenamefont {Lombard}, \citenamefont {Michau},
  \citenamefont {Hanna}, \citenamefont {Zaouter}, \citenamefont {Tajima},\ and\
  \citenamefont {Mourou}}]{brocklesby_ican_2014}%
  \BibitemOpen
  \bibfield  {author} {\bibinfo {author} {\bibfnamefont {W.~S.}\ \bibnamefont
  {Brocklesby}}, \bibinfo {author} {\bibfnamefont {J.}~\bibnamefont {Nilsson}},
  \bibinfo {author} {\bibfnamefont {T.}~\bibnamefont {Schreiber}}, \bibinfo
  {author} {\bibfnamefont {J.}~\bibnamefont {Limpert}}, \bibinfo {author}
  {\bibfnamefont {A.}~\bibnamefont {Brignon}}, \bibinfo {author} {\bibfnamefont
  {J.}~\bibnamefont {Bourderionnet}}, \bibinfo {author} {\bibfnamefont
  {L.}~\bibnamefont {Lombard}}, \bibinfo {author} {\bibfnamefont
  {V.}~\bibnamefont {Michau}}, \bibinfo {author} {\bibfnamefont
  {M.}~\bibnamefont {Hanna}}, \bibinfo {author} {\bibfnamefont
  {Y.}~\bibnamefont {Zaouter}}, \bibinfo {author} {\bibfnamefont
  {T.}~\bibnamefont {Tajima}}, and\ \bibinfo {author} {\bibfnamefont
  {G.}~\bibnamefont {Mourou}},\ }\bibfield  {title} {\bibinfo {title} {{ICAN}
  as a new laser paradigm for high energy, high average power femtosecond
  pulses},\ }\href {https://doi.org/10.1140/epjst/e2014-02172-4} {\ \textbf
  {\bibinfo {volume} {223}},\ \bibinfo {pages} {1189}}\BibitemShut {NoStop}%
\bibitem [{\citenamefont {Roso}(2018)}]{Roso2018high}%
  \BibitemOpen
  \bibfield  {author} {\bibinfo {author} {\bibfnamefont {L.}~\bibnamefont
  {Roso}},\ }\bibfield  {title} {\bibinfo {title} {High repetition rate
  petawatt lasers},\ }\href {https://doi.org/10.1051/epjconf/201816701001}
  {\bibfield  {journal} {\bibinfo  {journal} {EPJ Web of Conferences}\ }\textbf
  {\bibinfo {volume} {167}},\ \bibinfo {pages} {01001} (\bibinfo {year}
  {2018})}\BibitemShut {NoStop}%
\bibitem [{\citenamefont {Bayramian}\ \emph {et~al.}(2007)\citenamefont
  {Bayramian}, \citenamefont {Armstrong}, \citenamefont {Ault}, \citenamefont
  {Beach}, \citenamefont {Bibeau}, \citenamefont {Caird}, \citenamefont
  {Campbell}, \citenamefont {Chai}, \citenamefont {Dawson}, \citenamefont
  {Ebbers}, \citenamefont {Erlandson}, \citenamefont {Fei}, \citenamefont
  {Freitas}, \citenamefont {Kent}, \citenamefont {Liao}, \citenamefont
  {Ladran}, \citenamefont {Menapace}, \citenamefont {Molander}, \citenamefont
  {Payne}, \citenamefont {Peterson}, \citenamefont {Randles}, \citenamefont
  {Schaffers}, \citenamefont {Sutton}, \citenamefont {Tassano}, \citenamefont
  {Telford},\ and\ \citenamefont {Utterback}}]{Bayramian2007mercury}%
  \BibitemOpen
  \bibfield  {author} {\bibinfo {author} {\bibfnamefont {A.}~\bibnamefont
  {Bayramian}}, \bibinfo {author} {\bibfnamefont {P.}~\bibnamefont
  {Armstrong}}, \bibinfo {author} {\bibfnamefont {E.}~\bibnamefont {Ault}},
  \bibinfo {author} {\bibfnamefont {R.}~\bibnamefont {Beach}}, \bibinfo
  {author} {\bibfnamefont {C.}~\bibnamefont {Bibeau}}, \bibinfo {author}
  {\bibfnamefont {J.}~\bibnamefont {Caird}}, \bibinfo {author} {\bibfnamefont
  {R.}~\bibnamefont {Campbell}}, \bibinfo {author} {\bibfnamefont
  {B.}~\bibnamefont {Chai}}, \bibinfo {author} {\bibfnamefont {J.}~\bibnamefont
  {Dawson}}, \bibinfo {author} {\bibfnamefont {C.}~\bibnamefont {Ebbers}},
  \bibinfo {author} {\bibfnamefont {A.}~\bibnamefont {Erlandson}}, \bibinfo
  {author} {\bibfnamefont {Y.}~\bibnamefont {Fei}}, \bibinfo {author}
  {\bibfnamefont {B.}~\bibnamefont {Freitas}}, \bibinfo {author} {\bibfnamefont
  {R.}~\bibnamefont {Kent}}, \bibinfo {author} {\bibfnamefont {Z.}~\bibnamefont
  {Liao}}, \bibinfo {author} {\bibfnamefont {T.}~\bibnamefont {Ladran}},
  \bibinfo {author} {\bibfnamefont {J.}~\bibnamefont {Menapace}}, \bibinfo
  {author} {\bibfnamefont {B.}~\bibnamefont {Molander}}, \bibinfo {author}
  {\bibfnamefont {S.}~\bibnamefont {Payne}}, \bibinfo {author} {\bibfnamefont
  {N.}~\bibnamefont {Peterson}}, \bibinfo {author} {\bibfnamefont
  {M.}~\bibnamefont {Randles}}, \bibinfo {author} {\bibfnamefont
  {K.}~\bibnamefont {Schaffers}}, \bibinfo {author} {\bibfnamefont
  {S.}~\bibnamefont {Sutton}}, \bibinfo {author} {\bibfnamefont
  {J.}~\bibnamefont {Tassano}}, \bibinfo {author} {\bibfnamefont
  {S.}~\bibnamefont {Telford}}, and\ \bibinfo {author} {\bibfnamefont
  {E.}~\bibnamefont {Utterback}},\ }\bibfield  {title} {\bibinfo {title} {The
  mercury project: A high average power, gas-cooled laser for inertial fusion
  energy development},\ }\href {https://doi.org/10.13182/fst07-a1517}
  {\bibfield  {journal} {\bibinfo  {journal} {Fusion Science and Technology}\
  }\textbf {\bibinfo {volume} {52}},\ \bibinfo {pages} {383} (\bibinfo {year}
  {2007})}\BibitemShut {NoStop}%
\bibitem [{\citenamefont {Schaeffer}\ \emph {et~al.}(2018)\citenamefont
  {Schaeffer}, \citenamefont {Hofer}, \citenamefont {Knall}, \citenamefont
  {Heuer}, \citenamefont {Constantin},\ and\ \citenamefont
  {Niemann}}]{Schaeffer2018platform}%
  \BibitemOpen
  \bibfield  {author} {\bibinfo {author} {\bibfnamefont {D.~B.}\ \bibnamefont
  {Schaeffer}}, \bibinfo {author} {\bibfnamefont {L.~R.}\ \bibnamefont
  {Hofer}}, \bibinfo {author} {\bibfnamefont {E.~N.}\ \bibnamefont {Knall}},
  \bibinfo {author} {\bibfnamefont {P.~V.}\ \bibnamefont {Heuer}}, \bibinfo
  {author} {\bibfnamefont {C.~G.}\ \bibnamefont {Constantin}}, and\ \bibinfo
  {author} {\bibfnamefont {C.}~\bibnamefont {Niemann}},\ }\bibfield  {title}
  {\bibinfo {title} {A platform for high-repetition-rate laser experiments on
  the large plasma device},\ }\href {https://doi.org/10.1017/hpl.2018.11}
  {\bibfield  {journal} {\bibinfo  {journal} {High Power Laser Science and
  Engineering}\ }\textbf {\bibinfo {volume} {6}},\ \bibinfo {pages} {e17}
  (\bibinfo {year} {2018})}\BibitemShut {NoStop}%
\bibitem [{\citenamefont {Spears}\ \emph {et~al.}(2018)\citenamefont {Spears},
  \citenamefont {Brase}, \citenamefont {Bremer}, \citenamefont {Chen},
  \citenamefont {Field}, \citenamefont {Gaffney}, \citenamefont {Kruse},
  \citenamefont {Langer}, \citenamefont {Lewis}, \citenamefont {Nora},
  \citenamefont {Peterson}, \citenamefont {Thiagarajan}, \citenamefont
  {Essen},\ and\ \citenamefont {Humbird}}]{Spears2018deep}%
  \BibitemOpen
  \bibfield  {author} {\bibinfo {author} {\bibfnamefont {B.~K.}\ \bibnamefont
  {Spears}}, \bibinfo {author} {\bibfnamefont {J.}~\bibnamefont {Brase}},
  \bibinfo {author} {\bibfnamefont {P.-T.}\ \bibnamefont {Bremer}}, \bibinfo
  {author} {\bibfnamefont {B.}~\bibnamefont {Chen}}, \bibinfo {author}
  {\bibfnamefont {J.}~\bibnamefont {Field}}, \bibinfo {author} {\bibfnamefont
  {J.}~\bibnamefont {Gaffney}}, \bibinfo {author} {\bibfnamefont
  {M.}~\bibnamefont {Kruse}}, \bibinfo {author} {\bibfnamefont
  {S.}~\bibnamefont {Langer}}, \bibinfo {author} {\bibfnamefont
  {K.}~\bibnamefont {Lewis}}, \bibinfo {author} {\bibfnamefont
  {R.}~\bibnamefont {Nora}}, \bibinfo {author} {\bibfnamefont {J.~L.}\
  \bibnamefont {Peterson}}, \bibinfo {author} {\bibfnamefont {J.~J.}\
  \bibnamefont {Thiagarajan}}, \bibinfo {author} {\bibfnamefont {B.~V.}\
  \bibnamefont {Essen}}, and\ \bibinfo {author} {\bibfnamefont
  {K.}~\bibnamefont {Humbird}},\ }\bibfield  {title} {\bibinfo {title} {Deep
  learning: A guide for practitioners in the physical sciences},\ }\href
  {https://doi.org/10.1063/1.5020791} {\bibfield  {journal} {\bibinfo
  {journal} {Physics of Plasmas}\ }\textbf {\bibinfo {volume} {25}},\ \bibinfo
  {pages} {080901} (\bibinfo {year} {2018})}\BibitemShut {NoStop}%
\bibitem [{\citenamefont {Gopalaswamy}\ \emph {et~al.}(2019)\citenamefont
  {Gopalaswamy}, \citenamefont {Betti}, \citenamefont {Knauer}, \citenamefont
  {Luciani}, \citenamefont {Patel}, \citenamefont {Woo}, \citenamefont {Bose},
  \citenamefont {Igumenshchev}, \citenamefont {Campbell}, \citenamefont
  {Anderson}, \citenamefont {Bauer}, \citenamefont {Bonino}, \citenamefont
  {Cao}, \citenamefont {Christopherson}, \citenamefont {Collins}, \citenamefont
  {Collins}, \citenamefont {Davies}, \citenamefont {Delettrez}, \citenamefont
  {Edgell}, \citenamefont {Epstein}, \citenamefont {Forrest}, \citenamefont
  {Froula}, \citenamefont {Glebov}, \citenamefont {Goncharov}, \citenamefont
  {Harding}, \citenamefont {Hu}, \citenamefont {Jacobs-Perkins}, \citenamefont
  {Janezic}, \citenamefont {Kelly}, \citenamefont {Mannion}, \citenamefont
  {Maximov}, \citenamefont {Marshall}, \citenamefont {Michel}, \citenamefont
  {Miller}, \citenamefont {Morse}, \citenamefont {Palastro}, \citenamefont
  {Peebles}, \citenamefont {Radha}, \citenamefont {Regan}, \citenamefont
  {Sampat}, \citenamefont {Sangster}, \citenamefont {Sefkow}, \citenamefont
  {Seka}, \citenamefont {Shah}, \citenamefont {Shmyada}, \citenamefont
  {Shvydky}, \citenamefont {Stoeckl}, \citenamefont {Solodov}, \citenamefont
  {Theobald}, \citenamefont {Zuegel}, \citenamefont {Johnson}, \citenamefont
  {Petrasso}, \citenamefont {Li},\ and\ \citenamefont
  {Frenje}}]{Gopalaswamy2019tripled}%
  \BibitemOpen
  \bibfield  {author} {\bibinfo {author} {\bibfnamefont {V.}~\bibnamefont
  {Gopalaswamy}}, \bibinfo {author} {\bibfnamefont {R.}~\bibnamefont {Betti}},
  \bibinfo {author} {\bibfnamefont {J.~P.}\ \bibnamefont {Knauer}}, \bibinfo
  {author} {\bibfnamefont {N.}~\bibnamefont {Luciani}}, \bibinfo {author}
  {\bibfnamefont {D.}~\bibnamefont {Patel}}, \bibinfo {author} {\bibfnamefont
  {K.~M.}\ \bibnamefont {Woo}}, \bibinfo {author} {\bibfnamefont
  {A.}~\bibnamefont {Bose}}, \bibinfo {author} {\bibfnamefont {I.~V.}\
  \bibnamefont {Igumenshchev}}, \bibinfo {author} {\bibfnamefont {E.~M.}\
  \bibnamefont {Campbell}}, \bibinfo {author} {\bibfnamefont {K.~S.}\
  \bibnamefont {Anderson}}, \bibinfo {author} {\bibfnamefont {K.~A.}\
  \bibnamefont {Bauer}}, \bibinfo {author} {\bibfnamefont {M.~J.}\ \bibnamefont
  {Bonino}}, \bibinfo {author} {\bibfnamefont {D.}~\bibnamefont {Cao}},
  \bibinfo {author} {\bibfnamefont {A.~R.}\ \bibnamefont {Christopherson}},
  \bibinfo {author} {\bibfnamefont {G.~W.}\ \bibnamefont {Collins}}, \bibinfo
  {author} {\bibfnamefont {T.~J.~B.}\ \bibnamefont {Collins}}, \bibinfo
  {author} {\bibfnamefont {J.~R.}\ \bibnamefont {Davies}}, \bibinfo {author}
  {\bibfnamefont {J.~A.}\ \bibnamefont {Delettrez}}, \bibinfo {author}
  {\bibfnamefont {D.~H.}\ \bibnamefont {Edgell}}, \bibinfo {author}
  {\bibfnamefont {R.}~\bibnamefont {Epstein}}, \bibinfo {author} {\bibfnamefont
  {C.~J.}\ \bibnamefont {Forrest}}, \bibinfo {author} {\bibfnamefont {D.~H.}\
  \bibnamefont {Froula}}, \bibinfo {author} {\bibfnamefont {V.~Y.}\
  \bibnamefont {Glebov}}, \bibinfo {author} {\bibfnamefont {V.~N.}\
  \bibnamefont {Goncharov}}, \bibinfo {author} {\bibfnamefont {D.~R.}\
  \bibnamefont {Harding}}, \bibinfo {author} {\bibfnamefont {S.~X.}\
  \bibnamefont {Hu}}, \bibinfo {author} {\bibfnamefont {D.~W.}\ \bibnamefont
  {Jacobs-Perkins}}, \bibinfo {author} {\bibfnamefont {R.~T.}\ \bibnamefont
  {Janezic}}, \bibinfo {author} {\bibfnamefont {J.~H.}\ \bibnamefont {Kelly}},
  \bibinfo {author} {\bibfnamefont {O.~M.}\ \bibnamefont {Mannion}}, \bibinfo
  {author} {\bibfnamefont {A.}~\bibnamefont {Maximov}}, \bibinfo {author}
  {\bibfnamefont {F.~J.}\ \bibnamefont {Marshall}}, \bibinfo {author}
  {\bibfnamefont {D.~T.}\ \bibnamefont {Michel}}, \bibinfo {author}
  {\bibfnamefont {S.}~\bibnamefont {Miller}}, \bibinfo {author} {\bibfnamefont
  {S.~F.~B.}\ \bibnamefont {Morse}}, \bibinfo {author} {\bibfnamefont
  {J.}~\bibnamefont {Palastro}}, \bibinfo {author} {\bibfnamefont
  {J.}~\bibnamefont {Peebles}}, \bibinfo {author} {\bibfnamefont {P.~B.}\
  \bibnamefont {Radha}}, \bibinfo {author} {\bibfnamefont {S.~P.}\ \bibnamefont
  {Regan}}, \bibinfo {author} {\bibfnamefont {S.}~\bibnamefont {Sampat}},
  \bibinfo {author} {\bibfnamefont {T.~C.}\ \bibnamefont {Sangster}}, \bibinfo
  {author} {\bibfnamefont {A.~B.}\ \bibnamefont {Sefkow}}, \bibinfo {author}
  {\bibfnamefont {W.}~\bibnamefont {Seka}}, \bibinfo {author} {\bibfnamefont
  {R.~C.}\ \bibnamefont {Shah}}, \bibinfo {author} {\bibfnamefont {W.~T.}\
  \bibnamefont {Shmyada}}, \bibinfo {author} {\bibfnamefont {A.}~\bibnamefont
  {Shvydky}}, \bibinfo {author} {\bibfnamefont {C.}~\bibnamefont {Stoeckl}},
  \bibinfo {author} {\bibfnamefont {A.~A.}\ \bibnamefont {Solodov}}, \bibinfo
  {author} {\bibfnamefont {W.}~\bibnamefont {Theobald}}, \bibinfo {author}
  {\bibfnamefont {J.~D.}\ \bibnamefont {Zuegel}}, \bibinfo {author}
  {\bibfnamefont {M.~G.}\ \bibnamefont {Johnson}}, \bibinfo {author}
  {\bibfnamefont {R.~D.}\ \bibnamefont {Petrasso}}, \bibinfo {author}
  {\bibfnamefont {C.~K.}\ \bibnamefont {Li}}, and\ \bibinfo {author}
  {\bibfnamefont {J.~A.}\ \bibnamefont {Frenje}},\ }\bibfield  {title}
  {\bibinfo {title} {Tripled yield in direct-drive laser fusion through
  statistical modelling},\ }\href {https://doi.org/10.1038/s41586-019-0877-0}
  {\bibfield  {journal} {\bibinfo  {journal} {Nature}\ }\textbf {\bibinfo
  {volume} {565}},\ \bibinfo {pages} {581} (\bibinfo {year}
  {2019})}\BibitemShut {NoStop}%
\bibitem [{\citenamefont {Sinars}\ \emph {et~al.}()\citenamefont {Sinars},
  \citenamefont {Campbell}, \citenamefont {Cuneo}, \citenamefont {Jennings},
  \citenamefont {Peterson},\ and\ \citenamefont {Sefkow}}]{sinars_role_2015}%
  \BibitemOpen
  \bibfield  {author} {\bibinfo {author} {\bibfnamefont {D.~B.}\ \bibnamefont
  {Sinars}}, \bibinfo {author} {\bibfnamefont {E.~M.}\ \bibnamefont
  {Campbell}}, \bibinfo {author} {\bibfnamefont {M.~E.}\ \bibnamefont {Cuneo}},
  \bibinfo {author} {\bibfnamefont {C.~A.}\ \bibnamefont {Jennings}}, \bibinfo
  {author} {\bibfnamefont {K.~J.}\ \bibnamefont {Peterson}}, and\ \bibinfo
  {author} {\bibfnamefont {A.~B.}\ \bibnamefont {Sefkow}},\ }\bibfield  {title}
  {\bibinfo {title} {The role of magnetized liner inertial fusion as a pathway
  to fusion energy},\ }\href {https://doi.org/10.1007/s10894-015-0023-4} {\
  \textbf {\bibinfo {volume} {35}},\ \bibinfo {pages} {78}}\BibitemShut
  {NoStop}%
\bibitem [{\citenamefont {Malka}\ \emph {et~al.}()\citenamefont {Malka},
  \citenamefont {Faure}, \citenamefont {Gauduel}, \citenamefont {Lefebvre},
  \citenamefont {Rousse},\ and\ \citenamefont {Phuoc}}]{malka_principles_2008}%
  \BibitemOpen
  \bibfield  {author} {\bibinfo {author} {\bibfnamefont {V.}~\bibnamefont
  {Malka}}, \bibinfo {author} {\bibfnamefont {J.}~\bibnamefont {Faure}},
  \bibinfo {author} {\bibfnamefont {Y.~A.}\ \bibnamefont {Gauduel}}, \bibinfo
  {author} {\bibfnamefont {E.}~\bibnamefont {Lefebvre}}, \bibinfo {author}
  {\bibfnamefont {A.}~\bibnamefont {Rousse}}, and\ \bibinfo {author}
  {\bibfnamefont {K.~T.}\ \bibnamefont {Phuoc}},\ }\bibfield  {title} {\bibinfo
  {title} {Principles and applications of compact laser–plasma
  accelerators},\ }\href {https://doi.org/10.1038/nphys966} {\ \textbf
  {\bibinfo {volume} {4}},\ \bibinfo {pages} {447}}\BibitemShut {NoStop}%
\bibitem [{\citenamefont {Gales}\ and\ \citenamefont
  {Team}()}]{gales_laser_2015}%
  \BibitemOpen
  \bibfield  {author} {\bibinfo {author} {\bibfnamefont {S.}~\bibnamefont
  {Gales}}and\ \bibinfo {author} {\bibfnamefont {f.~t. E.-N.}\ \bibnamefont
  {Team}},\ }\bibfield  {title} {\bibinfo {title} {Laser driven nuclear science
  and applications: The need of high efficiency, high power and high repetition
  rate laser beams},\ }\href {https://doi.org/10.1140/epjst/e2015-02575-7} {\
  \textbf {\bibinfo {volume} {224}},\ \bibinfo {pages} {2631}}\BibitemShut
  {NoStop}%
\bibitem [{\citenamefont {Kitagawa}\ \emph {et~al.}()\citenamefont {Kitagawa},
  \citenamefont {Mori}, \citenamefont {Komeda}, \citenamefont {Ishii},
  \citenamefont {Hanayama}, \citenamefont {Fujita}, \citenamefont {Okihara},
  \citenamefont {Sekine}, \citenamefont {Sato}, \citenamefont {Kurita},
  \citenamefont {Kawashima}, \citenamefont {Kan}, \citenamefont {Nakamura},
  \citenamefont {Kondo}, \citenamefont {Fujine}, \citenamefont {Azuma},
  \citenamefont {Motohiro}, \citenamefont {Hioki}, \citenamefont {Kakeno},
  \citenamefont {Nishimura}, \citenamefont {Sunahara},\ and\ \citenamefont
  {Sentoku}}]{kitagawa_hi-rep._2013}%
  \BibitemOpen
  \bibfield  {author} {\bibinfo {author} {\bibfnamefont {Y.}~\bibnamefont
  {Kitagawa}}, \bibinfo {author} {\bibfnamefont {Y.}~\bibnamefont {Mori}},
  \bibinfo {author} {\bibfnamefont {O.}~\bibnamefont {Komeda}}, \bibinfo
  {author} {\bibfnamefont {K.}~\bibnamefont {Ishii}}, \bibinfo {author}
  {\bibfnamefont {R.}~\bibnamefont {Hanayama}}, \bibinfo {author}
  {\bibfnamefont {K.}~\bibnamefont {Fujita}}, \bibinfo {author} {\bibfnamefont
  {S.-i.}\ \bibnamefont {Okihara}}, \bibinfo {author} {\bibfnamefont
  {T.}~\bibnamefont {Sekine}}, \bibinfo {author} {\bibfnamefont
  {N.}~\bibnamefont {Sato}}, \bibinfo {author} {\bibfnamefont {T.}~\bibnamefont
  {Kurita}}, \bibinfo {author} {\bibfnamefont {T.}~\bibnamefont {Kawashima}},
  \bibinfo {author} {\bibfnamefont {H.}~\bibnamefont {Kan}}, \bibinfo {author}
  {\bibfnamefont {N.}~\bibnamefont {Nakamura}}, \bibinfo {author}
  {\bibfnamefont {T.}~\bibnamefont {Kondo}}, \bibinfo {author} {\bibfnamefont
  {M.}~\bibnamefont {Fujine}}, \bibinfo {author} {\bibfnamefont
  {H.}~\bibnamefont {Azuma}}, \bibinfo {author} {\bibfnamefont
  {T.}~\bibnamefont {Motohiro}}, \bibinfo {author} {\bibfnamefont
  {T.}~\bibnamefont {Hioki}}, \bibinfo {author} {\bibfnamefont
  {M.}~\bibnamefont {Kakeno}}, \bibinfo {author} {\bibfnamefont
  {Y.}~\bibnamefont {Nishimura}}, \bibinfo {author} {\bibfnamefont
  {A.}~\bibnamefont {Sunahara}}, and\ \bibinfo {author} {\bibfnamefont
  {Y.}~\bibnamefont {Sentoku}},\ }\bibfield  {title} {\bibinfo {title} {Hi-rep.
  counter-illumination fast ignition scheme fusion},\ }\href
  {https://doi.org/10.1585/pfr.8.3404047} {\ \textbf {\bibinfo {volume} {8}},\
  \bibinfo {pages} {3404047}}\BibitemShut {NoStop}%
\bibitem [{\citenamefont {Masood}\ \emph {et~al.}()\citenamefont {Masood},
  \citenamefont {Baumann}, \citenamefont {Cowan}, \citenamefont {Enghardt},
  \citenamefont {Herrmannsdoerfer}, \citenamefont {Hofmann}, \citenamefont
  {Karsch}, \citenamefont {Kroll}, \citenamefont {Pawelke}, \citenamefont
  {Schramm},\ and\ \citenamefont {{others}}}]{masood2016}%
  \BibitemOpen
  \bibfield  {author} {\bibinfo {author} {\bibfnamefont {U.}~\bibnamefont
  {Masood}}, \bibinfo {author} {\bibfnamefont {M.}~\bibnamefont {Baumann}},
  \bibinfo {author} {\bibfnamefont {T.}~\bibnamefont {Cowan}}, \bibinfo
  {author} {\bibfnamefont {W.}~\bibnamefont {Enghardt}}, \bibinfo {author}
  {\bibfnamefont {T.}~\bibnamefont {Herrmannsdoerfer}}, \bibinfo {author}
  {\bibfnamefont {K.}~\bibnamefont {Hofmann}}, \bibinfo {author} {\bibfnamefont
  {L.}~\bibnamefont {Karsch}}, \bibinfo {author} {\bibfnamefont
  {F.}~\bibnamefont {Kroll}}, \bibinfo {author} {\bibfnamefont
  {J.}~\bibnamefont {Pawelke}}, \bibinfo {author} {\bibfnamefont
  {U.}~\bibnamefont {Schramm}}, and\ \bibinfo {author} {\bibnamefont
  {{others}}},\ }\bibfield  {title} {\bibinfo {title} {Novel approach to
  utilize proton beams from high power laser accelerators for therapy},\ }in\
  \href {http://www.jacow.org/ipac2016/papers/tupoy003.pdf} {\emph {\bibinfo
  {booktitle} {7th International Particle Accelerator Conference ({IPAC}'16),
  Busan, Korea, May 8-13, 2016}}}\ (\bibinfo  {publisher} {{JACOW}, Geneva,
  Switzerland})\ pp.\ \bibinfo {pages} {1905--1907}\BibitemShut {NoStop}%
\bibitem [{NAS()}]{NAS_2019}%
  \BibitemOpen
  \href {https://doi.org/https://doi.org.10.17226/24939} {\emph {\bibinfo
  {title} {National Academies of Sciences, Engineering, and Medicine}}}\
  (\bibinfo  {publisher} {The National Academies Press},\ \bibinfo {address}
  {Washington, D.C.})\BibitemShut {NoStop}%
\bibitem [{\citenamefont {Mourou}\ \emph {et~al.}()\citenamefont {Mourou},
  \citenamefont {Korn}, \citenamefont {Sandner},\ and\ \citenamefont
  {Collier}}]{ELI_whitebook}%
  \BibitemOpen
  \bibinfo {editor} {\bibfnamefont {G.}~\bibnamefont {Mourou}}, \bibinfo
  {editor} {\bibfnamefont {G.}~\bibnamefont {Korn}}, \bibinfo {editor}
  {\bibfnamefont {W.}~\bibnamefont {Sandner}}, and\ \bibinfo {editor}
  {\bibfnamefont {J.}~\bibnamefont {Collier}},\ eds.,\ \href
  {http://www.eli-np.ro/documents/ELI-NP-WhiteBook.pdf} {\emph {\bibinfo
  {title} {Extreme Light Infrastructure Whitebook: Science and Technology with
  Ultra-Intense Lasers}}}\ (\bibinfo  {publisher} {Thoss Media GmbH},\ \bibinfo
  {address} {Berlin})\BibitemShut {NoStop}%
\bibitem [{\citenamefont {Prencipe}\ \emph {et~al.}(2017)\citenamefont
  {Prencipe}, \citenamefont {Fuchs}, \citenamefont {Pascarelli}, \citenamefont
  {Schumacher}, \citenamefont {Stephens}, \citenamefont {Alexander},
  \citenamefont {Briggs}, \citenamefont {Büscher}, \citenamefont {Cernaianu},
  \citenamefont {Choukourov}, \citenamefont {Marco}, \citenamefont {Erbe},
  \citenamefont {Fassbender}, \citenamefont {Fiquet}, \citenamefont
  {Fitzsimmons}, \citenamefont {Gheorghiu}, \citenamefont {Hund}, \citenamefont
  {Huang}, \citenamefont {Harmand}, \citenamefont {Hartley}, \citenamefont
  {Irman}, \citenamefont {Kluge}, \citenamefont {Konopkova}, \citenamefont
  {Kraft}, \citenamefont {Kraus}, \citenamefont {Leca}, \citenamefont
  {Margarone}, \citenamefont {Metzkes}, \citenamefont {Nagai}, \citenamefont
  {Nazarov}, \citenamefont {Lutoslawski}, \citenamefont {Papp}, \citenamefont
  {Passoni}, \citenamefont {Pelka}, \citenamefont {Perin}, \citenamefont
  {Schulz}, \citenamefont {Smid}, \citenamefont {Spindloe}, \citenamefont
  {Steinke}, \citenamefont {Torchio}, \citenamefont {Vass}, \citenamefont
  {Wiste}, \citenamefont {Zaffino}, \citenamefont {Zeil}, \citenamefont
  {Tschentscher}, \citenamefont {Schramm},\ and\ \citenamefont
  {Cowan}}]{Prencipe2017targets}%
  \BibitemOpen
  \bibfield  {author} {\bibinfo {author} {\bibfnamefont {I.}~\bibnamefont
  {Prencipe}}, \bibinfo {author} {\bibfnamefont {J.}~\bibnamefont {Fuchs}},
  \bibinfo {author} {\bibfnamefont {S.}~\bibnamefont {Pascarelli}}, \bibinfo
  {author} {\bibfnamefont {D.~W.}\ \bibnamefont {Schumacher}}, \bibinfo
  {author} {\bibfnamefont {R.~B.}\ \bibnamefont {Stephens}}, \bibinfo {author}
  {\bibfnamefont {N.~B.}\ \bibnamefont {Alexander}}, \bibinfo {author}
  {\bibfnamefont {R.}~\bibnamefont {Briggs}}, \bibinfo {author} {\bibfnamefont
  {M.}~\bibnamefont {Büscher}}, \bibinfo {author} {\bibfnamefont {M.~O.}\
  \bibnamefont {Cernaianu}}, \bibinfo {author} {\bibfnamefont {A.}~\bibnamefont
  {Choukourov}}, \bibinfo {author} {\bibfnamefont {M.~D.}\ \bibnamefont
  {Marco}}, \bibinfo {author} {\bibfnamefont {A.}~\bibnamefont {Erbe}},
  \bibinfo {author} {\bibfnamefont {J.}~\bibnamefont {Fassbender}}, \bibinfo
  {author} {\bibfnamefont {G.}~\bibnamefont {Fiquet}}, \bibinfo {author}
  {\bibfnamefont {P.}~\bibnamefont {Fitzsimmons}}, \bibinfo {author}
  {\bibfnamefont {C.}~\bibnamefont {Gheorghiu}}, \bibinfo {author}
  {\bibfnamefont {J.}~\bibnamefont {Hund}}, \bibinfo {author} {\bibfnamefont
  {L.~G.}\ \bibnamefont {Huang}}, \bibinfo {author} {\bibfnamefont
  {M.}~\bibnamefont {Harmand}}, \bibinfo {author} {\bibfnamefont {N.~J.}\
  \bibnamefont {Hartley}}, \bibinfo {author} {\bibfnamefont {A.}~\bibnamefont
  {Irman}}, \bibinfo {author} {\bibfnamefont {T.}~\bibnamefont {Kluge}},
  \bibinfo {author} {\bibfnamefont {Z.}~\bibnamefont {Konopkova}}, \bibinfo
  {author} {\bibfnamefont {S.}~\bibnamefont {Kraft}}, \bibinfo {author}
  {\bibfnamefont {D.}~\bibnamefont {Kraus}}, \bibinfo {author} {\bibfnamefont
  {V.}~\bibnamefont {Leca}}, \bibinfo {author} {\bibfnamefont {D.}~\bibnamefont
  {Margarone}}, \bibinfo {author} {\bibfnamefont {J.}~\bibnamefont {Metzkes}},
  \bibinfo {author} {\bibfnamefont {K.}~\bibnamefont {Nagai}}, \bibinfo
  {author} {\bibfnamefont {W.}~\bibnamefont {Nazarov}}, \bibinfo {author}
  {\bibfnamefont {P.}~\bibnamefont {Lutoslawski}}, \bibinfo {author}
  {\bibfnamefont {D.}~\bibnamefont {Papp}}, \bibinfo {author} {\bibfnamefont
  {M.}~\bibnamefont {Passoni}}, \bibinfo {author} {\bibfnamefont
  {A.}~\bibnamefont {Pelka}}, \bibinfo {author} {\bibfnamefont {J.~P.}\
  \bibnamefont {Perin}}, \bibinfo {author} {\bibfnamefont {J.}~\bibnamefont
  {Schulz}}, \bibinfo {author} {\bibfnamefont {M.}~\bibnamefont {Smid}},
  \bibinfo {author} {\bibfnamefont {C.}~\bibnamefont {Spindloe}}, \bibinfo
  {author} {\bibfnamefont {S.}~\bibnamefont {Steinke}}, \bibinfo {author}
  {\bibfnamefont {R.}~\bibnamefont {Torchio}}, \bibinfo {author} {\bibfnamefont
  {C.}~\bibnamefont {Vass}}, \bibinfo {author} {\bibfnamefont {T.}~\bibnamefont
  {Wiste}}, \bibinfo {author} {\bibfnamefont {R.}~\bibnamefont {Zaffino}},
  \bibinfo {author} {\bibfnamefont {K.}~\bibnamefont {Zeil}}, \bibinfo {author}
  {\bibfnamefont {T.}~\bibnamefont {Tschentscher}}, \bibinfo {author}
  {\bibfnamefont {U.}~\bibnamefont {Schramm}}, and\ \bibinfo {author}
  {\bibfnamefont {T.~E.}\ \bibnamefont {Cowan}},\ }\bibfield  {title} {\bibinfo
  {title} {Targets for high repetition rate laser facilities: needs, challenges
  and perspectives},\ }\bibfield  {journal} {\bibinfo  {journal} {High Power
  Laser Science and Engineering}\ }\textbf {\bibinfo {volume} {5}},\ \href
  {https://doi.org/10.1017/hpl.2017.18} {10.1017/hpl.2017.18} (\bibinfo {year}
  {2017})\BibitemShut {NoStop}%
\bibitem [{\citenamefont {Grim}\ \emph {et~al.}(2018)\citenamefont {Grim},
  \citenamefont {Kemp}, \citenamefont {Wilks}, \citenamefont {Hartouni},\ and\
  \citenamefont {Kerr}}]{grim_foam_2018}%
  \BibitemOpen
  \bibfield  {author} {\bibinfo {author} {\bibfnamefont {G.}~\bibnamefont
  {Grim}}, \bibinfo {author} {\bibfnamefont {A.~J.}\ \bibnamefont {Kemp}},
  \bibinfo {author} {\bibfnamefont {S.}~\bibnamefont {Wilks}}, \bibinfo
  {author} {\bibfnamefont {E.~P.}\ \bibnamefont {Hartouni}}, and\ \bibinfo
  {author} {\bibfnamefont {S.}~\bibnamefont {Kerr}},\ }\bibfield  {title}
  {\bibinfo {title} {Generating near solid density reacting ion distributions
  using intense short pulse lasers},\ }\href
  {http://meetings.aps.org/link/BAPS.2018.DPP.TO6.8} {\bibfield  {journal}
  {\bibinfo  {journal} {Bulletin of the APS DPP meeting 2018}\ } (\bibinfo
  {year} {2018})}\BibitemShut {NoStop}%
\bibitem [{\citenamefont {et~al.}()}]{poole_LCdamage_2019}%
  \BibitemOpen
  \bibfield  {author} {\bibinfo {author} {\bibfnamefont {P.~L.~P.}\
  \bibnamefont {et~al.}},\ }\bibinfo {title} {in preparation}\BibitemShut
  {NoStop}%
\bibitem [{\citenamefont {Lübcke}\ \emph {et~al.}()\citenamefont {Lübcke},
  \citenamefont {Andreev}, \citenamefont {Höhm}, \citenamefont {Grunwald},
  \citenamefont {Ehrentraut},\ and\ \citenamefont
  {Schnürer}}]{lubcke_prospects_2017}%
  \BibitemOpen
\bibfield  {title} {  }\bibfield  {author} {\bibinfo {author} {\bibfnamefont
  {A.}~\bibnamefont {Lübcke}}, \bibinfo {author} {\bibfnamefont {A.~A.}\
  \bibnamefont {Andreev}}, \bibinfo {author} {\bibfnamefont {S.}~\bibnamefont
  {Höhm}}, \bibinfo {author} {\bibfnamefont {R.}~\bibnamefont {Grunwald}},
  \bibinfo {author} {\bibfnamefont {L.}~\bibnamefont {Ehrentraut}}, and\
  \bibinfo {author} {\bibfnamefont {M.}~\bibnamefont {Schnürer}},\ }\bibfield
  {title} {\bibinfo {title} {Prospects of target nanostructuring for laser
  proton acceleration}\ }\textbf {\bibinfo {volume} {7}},\ \href
  {https://doi.org/10.1038/srep44030} {10.1038/srep44030}\BibitemShut {NoStop}%
\bibitem [{\citenamefont {Grittani}\ \emph {et~al.}()\citenamefont {Grittani},
  \citenamefont {Anania}, \citenamefont {Gatti}, \citenamefont {Giulietti},
  \citenamefont {Kando}, \citenamefont {Krus}, \citenamefont {Labate},
  \citenamefont {Levato}, \citenamefont {Londrillo}, \citenamefont {Rossi},\
  and\ \citenamefont {Gizzi}}]{grittani_high_2014}%
  \BibitemOpen
  \bibfield  {author} {\bibinfo {author} {\bibfnamefont {G.}~\bibnamefont
  {Grittani}}, \bibinfo {author} {\bibfnamefont {M.~P.}\ \bibnamefont
  {Anania}}, \bibinfo {author} {\bibfnamefont {G.}~\bibnamefont {Gatti}},
  \bibinfo {author} {\bibfnamefont {D.}~\bibnamefont {Giulietti}}, \bibinfo
  {author} {\bibfnamefont {M.}~\bibnamefont {Kando}}, \bibinfo {author}
  {\bibfnamefont {M.}~\bibnamefont {Krus}}, \bibinfo {author} {\bibfnamefont
  {L.}~\bibnamefont {Labate}}, \bibinfo {author} {\bibfnamefont
  {T.}~\bibnamefont {Levato}}, \bibinfo {author} {\bibfnamefont
  {P.}~\bibnamefont {Londrillo}}, \bibinfo {author} {\bibfnamefont
  {F.}~\bibnamefont {Rossi}}, and\ \bibinfo {author} {\bibfnamefont {L.~A.}\
  \bibnamefont {Gizzi}},\ }\bibfield  {title} {\bibinfo {title} {High energy
  electrons from interaction with a structured gas-jet at {FLAME}},\ }\href
  {https://doi.org/10.1016/j.nima.2013.10.082} {\ \bibinfo {series}
  {Proceedings of the first European Advanced Accelerator Concepts Workshop
  2013},\ \textbf {\bibinfo {volume} {740}},\ \bibinfo {pages}
  {257}}\BibitemShut {NoStop}%
\bibitem [{\citenamefont {Hoath}\ \emph {et~al.}()\citenamefont {Hoath},
  \citenamefont {Martin},\ and\ \citenamefont
  {Hutchings}}]{hoath_atomization_2010}%
  \BibitemOpen
  \bibfield  {author} {\bibinfo {author} {\bibfnamefont {S.~D.}\ \bibnamefont
  {Hoath}}, \bibinfo {author} {\bibfnamefont {G.~D.}\ \bibnamefont {Martin}},
  and\ \bibinfo {author} {\bibfnamefont {I.~M.}\ \bibnamefont {Hutchings}},\
  }\bibfield  {title} {\bibinfo {title} {Atomization patterns produced by the
  oblique collision of two newtonian liquid jets},\ }\href
  {https://doi.org/10.1063/1.3373513} {\ \textbf {\bibinfo {volume} {22}},\
  \bibinfo {pages} {042101}}\BibitemShut {NoStop}%
\bibitem [{\citenamefont {Poole}\ \emph
  {et~al.}(2016{\natexlab{a}})\citenamefont {Poole}, \citenamefont {Krygier},
  \citenamefont {Cochran}, \citenamefont {Foster}, \citenamefont {Scott},
  \citenamefont {Wilson}, \citenamefont {Bailey}, \citenamefont {Bourgeois},
  \citenamefont {Hernandez-Gomez}, \citenamefont {Neely}, \citenamefont
  {Rajeev}, \citenamefont {Freeman},\ and\ \citenamefont
  {Schumacher}}]{poole_plasma_2016}%
  \BibitemOpen
  \bibfield  {author} {\bibinfo {author} {\bibfnamefont {P.~L.}\ \bibnamefont
  {Poole}}, \bibinfo {author} {\bibfnamefont {A.}~\bibnamefont {Krygier}},
  \bibinfo {author} {\bibfnamefont {G.~E.}\ \bibnamefont {Cochran}}, \bibinfo
  {author} {\bibfnamefont {P.~S.}\ \bibnamefont {Foster}}, \bibinfo {author}
  {\bibfnamefont {G.~G.}\ \bibnamefont {Scott}}, \bibinfo {author}
  {\bibfnamefont {L.~A.}\ \bibnamefont {Wilson}}, \bibinfo {author}
  {\bibfnamefont {J.}~\bibnamefont {Bailey}}, \bibinfo {author} {\bibfnamefont
  {N.}~\bibnamefont {Bourgeois}}, \bibinfo {author} {\bibfnamefont
  {C.}~\bibnamefont {Hernandez-Gomez}}, \bibinfo {author} {\bibfnamefont
  {D.}~\bibnamefont {Neely}}, \bibinfo {author} {\bibfnamefont {P.~P.}\
  \bibnamefont {Rajeev}}, \bibinfo {author} {\bibfnamefont {R.~R.}\
  \bibnamefont {Freeman}}, and\ \bibinfo {author} {\bibfnamefont {D.~W.}\
  \bibnamefont {Schumacher}},\ }\bibfield  {title} {\bibinfo {title}
  {Experiment and simulation of novel liquid crystal plasma mirrors for high
  contrast, intense laser pulses},\ }\href {https://doi.org/10.1038/srep32041}
  {\bibfield  {journal} {\bibinfo  {journal} {Scientific Reports}\ }\textbf
  {\bibinfo {volume} {6}},\ \bibinfo {pages} {32041} (\bibinfo {year}
  {2016}{\natexlab{a}})}\BibitemShut {NoStop}%
\bibitem [{\citenamefont {Shaw}\ \emph {et~al.}(2016)\citenamefont {Shaw},
  \citenamefont {Steinke}, \citenamefont {van Tilborg},\ and\ \citenamefont
  {Leemans}}]{shaw_tape_2016}%
  \BibitemOpen
  \bibfield  {author} {\bibinfo {author} {\bibfnamefont {B.~H.}\ \bibnamefont
  {Shaw}}, \bibinfo {author} {\bibfnamefont {S.}~\bibnamefont {Steinke}},
  \bibinfo {author} {\bibfnamefont {J.}~\bibnamefont {van Tilborg}}, and\
  \bibinfo {author} {\bibfnamefont {W.~P.}\ \bibnamefont {Leemans}},\
  }\bibfield  {title} {\bibinfo {title} {Reflectance characterization of
  tape-based plasma mirrors},\ }\href {https://doi.org/10.1063/1.4954242}
  {\bibfield  {journal} {\bibinfo  {journal} {Physics of Plasmas}\ }\textbf
  {\bibinfo {volume} {23}},\ \bibinfo {pages} {063118} (\bibinfo {year}
  {2016})},\ \Eprint {https://arxiv.org/abs/https://doi.org/10.1063/1.4954242}
  {https://doi.org/10.1063/1.4954242} \BibitemShut {NoStop}%
\bibitem [{\citenamefont {Salehi}\ \emph {et~al.}()\citenamefont {Salehi},
  \citenamefont {Goers}, \citenamefont {Hine}, \citenamefont {Feder},
  \citenamefont {Kuk}, \citenamefont {Miao}, \citenamefont {Woodbury},
  \citenamefont {Kim},\ and\ \citenamefont {Milchberg}}]{salehi_mev_2017}%
  \BibitemOpen
  \bibfield  {author} {\bibinfo {author} {\bibfnamefont {F.}~\bibnamefont
  {Salehi}}, \bibinfo {author} {\bibfnamefont {A.~J.}\ \bibnamefont {Goers}},
  \bibinfo {author} {\bibfnamefont {G.~A.}\ \bibnamefont {Hine}}, \bibinfo
  {author} {\bibfnamefont {L.}~\bibnamefont {Feder}}, \bibinfo {author}
  {\bibfnamefont {D.}~\bibnamefont {Kuk}}, \bibinfo {author} {\bibfnamefont
  {B.}~\bibnamefont {Miao}}, \bibinfo {author} {\bibfnamefont {D.}~\bibnamefont
  {Woodbury}}, \bibinfo {author} {\bibfnamefont {K.~Y.}\ \bibnamefont {Kim}},
  and\ \bibinfo {author} {\bibfnamefont {H.~M.}\ \bibnamefont {Milchberg}},\
  }\bibfield  {title} {\bibinfo {title} {{MeV} electron acceleration at 1 {kHz}
  with {\textless}10 {mJ} laser pulses},\ }\href
  {https://doi.org/10.1364/OL.42.000215} {\ \textbf {\bibinfo {volume} {42}},\
  \bibinfo {pages} {215}}\BibitemShut {NoStop}%
\bibitem [{\citenamefont {Aniculaesei}\ \emph {et~al.}(2018)\citenamefont
  {Aniculaesei}, \citenamefont {Kim}, \citenamefont {Yoo}, \citenamefont {Oh},\
  and\ \citenamefont {Nam}}]{Aniculaesei2018novel}%
  \BibitemOpen
  \bibfield  {author} {\bibinfo {author} {\bibfnamefont {C.}~\bibnamefont
  {Aniculaesei}}, \bibinfo {author} {\bibfnamefont {H.~T.}\ \bibnamefont
  {Kim}}, \bibinfo {author} {\bibfnamefont {B.~J.}\ \bibnamefont {Yoo}},
  \bibinfo {author} {\bibfnamefont {K.~H.}\ \bibnamefont {Oh}}, and\ \bibinfo
  {author} {\bibfnamefont {C.~H.}\ \bibnamefont {Nam}},\ }\bibfield  {title}
  {\bibinfo {title} {Novel gas target for laser wakefield accelerators},\
  }\href {https://doi.org/10.1063/1.4993269} {\bibfield  {journal} {\bibinfo
  {journal} {Review of Scientific Instruments}\ }\textbf {\bibinfo {volume}
  {89}},\ \bibinfo {pages} {025110} (\bibinfo {year} {2018})}\BibitemShut
  {NoStop}%
\bibitem [{\citenamefont {Lorenz}\ \emph {et~al.}(2019)\citenamefont {Lorenz},
  \citenamefont {Grittani}, \citenamefont {Chacon-Golcher}, \citenamefont
  {Lazzarini}, \citenamefont {Limpouch}, \citenamefont {Nawaz}, \citenamefont
  {Nevrkla}, \citenamefont {Vilanova},\ and\ \citenamefont
  {Levato}}]{Lorenz2019characterization}%
  \BibitemOpen
  \bibfield  {author} {\bibinfo {author} {\bibfnamefont {S.}~\bibnamefont
  {Lorenz}}, \bibinfo {author} {\bibfnamefont {G.}~\bibnamefont {Grittani}},
  \bibinfo {author} {\bibfnamefont {E.}~\bibnamefont {Chacon-Golcher}},
  \bibinfo {author} {\bibfnamefont {C.~M.}\ \bibnamefont {Lazzarini}}, \bibinfo
  {author} {\bibfnamefont {J.}~\bibnamefont {Limpouch}}, \bibinfo {author}
  {\bibfnamefont {F.}~\bibnamefont {Nawaz}}, \bibinfo {author} {\bibfnamefont
  {M.}~\bibnamefont {Nevrkla}}, \bibinfo {author} {\bibfnamefont
  {L.}~\bibnamefont {Vilanova}}, and\ \bibinfo {author} {\bibfnamefont
  {T.}~\bibnamefont {Levato}},\ }\bibfield  {title} {\bibinfo {title}
  {Characterization of supersonic and subsonic gas targets for laser wakefield
  electron acceleration experiments},\ }\href
  {https://doi.org/10.1063/1.5081509} {\bibfield  {journal} {\bibinfo
  {journal} {Matter and Radiation at Extremes}\ }\textbf {\bibinfo {volume}
  {4}},\ \bibinfo {pages} {015401} (\bibinfo {year} {2019})}\BibitemShut
  {NoStop}%
\bibitem [{\citenamefont {George}\ \emph {et~al.}(2019)\citenamefont {George},
  \citenamefont {Morrison}, \citenamefont {Feister}, \citenamefont {Ngirmang},
  \citenamefont {Smith}, \citenamefont {Klim}, \citenamefont {Snyder},
  \citenamefont {Austin}, \citenamefont {Erbsen}, \citenamefont {Frische} \emph
  {et~al.}}]{george_high_2019}%
  \BibitemOpen
  \bibfield  {author} {\bibinfo {author} {\bibfnamefont {K.}~\bibnamefont
  {George}}, \bibinfo {author} {\bibfnamefont {J.}~\bibnamefont {Morrison}},
  \bibinfo {author} {\bibfnamefont {S.}~\bibnamefont {Feister}}, \bibinfo
  {author} {\bibfnamefont {G.}~\bibnamefont {Ngirmang}}, \bibinfo {author}
  {\bibfnamefont {J.}~\bibnamefont {Smith}}, \bibinfo {author} {\bibfnamefont
  {A.}~\bibnamefont {Klim}}, \bibinfo {author} {\bibfnamefont {J.}~\bibnamefont
  {Snyder}}, \bibinfo {author} {\bibfnamefont {D.}~\bibnamefont {Austin}},
  \bibinfo {author} {\bibfnamefont {W.}~\bibnamefont {Erbsen}}, \bibinfo
  {author} {\bibfnamefont {K.}~\bibnamefont {Frische}},  \emph {et~al.},\
  }\bibfield  {title} {\bibinfo {title} {High repetition rate (>= khz) targets
  and optics from liquid microjets for the study and application of high
  intensity laser-plasma interactions},\ }\href
  {https://arxiv.org/abs/1902.04656} {\bibfield  {journal} {\bibinfo  {journal}
  {arXiv preprint arXiv:1902.04656}\ } (\bibinfo {year} {2019})}\BibitemShut
  {NoStop}%
\bibitem [{\citenamefont {Schwind}\ \emph {et~al.}(2019)\citenamefont
  {Schwind}, \citenamefont {Aktan}, \citenamefont {Cerchez}, \citenamefont
  {Prasad}, \citenamefont {Willi},\ and\ \citenamefont
  {Aurand}}]{Schwind2019high}%
  \BibitemOpen
  \bibfield  {author} {\bibinfo {author} {\bibfnamefont {K.}~\bibnamefont
  {Schwind}}, \bibinfo {author} {\bibfnamefont {E.}~\bibnamefont {Aktan}},
  \bibinfo {author} {\bibfnamefont {M.}~\bibnamefont {Cerchez}}, \bibinfo
  {author} {\bibfnamefont {R.}~\bibnamefont {Prasad}}, \bibinfo {author}
  {\bibfnamefont {O.}~\bibnamefont {Willi}}, and\ \bibinfo {author}
  {\bibfnamefont {B.}~\bibnamefont {Aurand}},\ }\bibfield  {title} {\bibinfo
  {title} {A high-repetition rate droplet-source for plasma physics
  applications},\ }\href {https://doi.org/10.1016/j.nima.2019.03.004}
  {\bibfield  {journal} {\bibinfo  {journal} {Nuclear Instruments and Methods
  in Physics Research Section A: Accelerators, Spectrometers, Detectors and
  Associated Equipment}\ }\textbf {\bibinfo {volume} {928}},\ \bibinfo {pages}
  {65} (\bibinfo {year} {2019})}\BibitemShut {NoStop}%
\bibitem [{\citenamefont {Stan}\ \emph {et~al.}()\citenamefont {Stan},
  \citenamefont {Milathianaki}, \citenamefont {Laksmono}, \citenamefont
  {Sierra}, \citenamefont {{McQueen}}, \citenamefont {Messerschmidt},
  \citenamefont {Williams}, \citenamefont {Koglin}, \citenamefont {Lane},
  \citenamefont {Hayes}, \citenamefont {Guillet}, \citenamefont {Liang},
  \citenamefont {Aquila}, \citenamefont {Willmott}, \citenamefont {Robinson},
  \citenamefont {Gumerlock}, \citenamefont {Botha}, \citenamefont {Nass},
  \citenamefont {Schlichting}, \citenamefont {Shoeman}, \citenamefont {Stone},\
  and\ \citenamefont {Boutet}}]{stan_liquid_2016}%
  \BibitemOpen
  \bibfield  {author} {\bibinfo {author} {\bibfnamefont {C.~A.}\ \bibnamefont
  {Stan}}, \bibinfo {author} {\bibfnamefont {D.}~\bibnamefont {Milathianaki}},
  \bibinfo {author} {\bibfnamefont {H.}~\bibnamefont {Laksmono}}, \bibinfo
  {author} {\bibfnamefont {R.~G.}\ \bibnamefont {Sierra}}, \bibinfo {author}
  {\bibfnamefont {T.~A.}\ \bibnamefont {{McQueen}}}, \bibinfo {author}
  {\bibfnamefont {M.}~\bibnamefont {Messerschmidt}}, \bibinfo {author}
  {\bibfnamefont {G.~J.}\ \bibnamefont {Williams}}, \bibinfo {author}
  {\bibfnamefont {J.~E.}\ \bibnamefont {Koglin}}, \bibinfo {author}
  {\bibfnamefont {T.~J.}\ \bibnamefont {Lane}}, \bibinfo {author}
  {\bibfnamefont {M.~J.}\ \bibnamefont {Hayes}}, \bibinfo {author}
  {\bibfnamefont {S.~A.~H.}\ \bibnamefont {Guillet}}, \bibinfo {author}
  {\bibfnamefont {M.}~\bibnamefont {Liang}}, \bibinfo {author} {\bibfnamefont
  {A.~L.}\ \bibnamefont {Aquila}}, \bibinfo {author} {\bibfnamefont {P.~R.}\
  \bibnamefont {Willmott}}, \bibinfo {author} {\bibfnamefont {J.~S.}\
  \bibnamefont {Robinson}}, \bibinfo {author} {\bibfnamefont {K.~L.}\
  \bibnamefont {Gumerlock}}, \bibinfo {author} {\bibfnamefont {S.}~\bibnamefont
  {Botha}}, \bibinfo {author} {\bibfnamefont {K.}~\bibnamefont {Nass}},
  \bibinfo {author} {\bibfnamefont {I.}~\bibnamefont {Schlichting}}, \bibinfo
  {author} {\bibfnamefont {R.~L.}\ \bibnamefont {Shoeman}}, \bibinfo {author}
  {\bibfnamefont {H.~A.}\ \bibnamefont {Stone}}, and\ \bibinfo {author}
  {\bibfnamefont {S.}~\bibnamefont {Boutet}},\ }\bibfield  {title} {\bibinfo
  {title} {Liquid explosions induced by x-ray laser pulses}\ }\textbf {\bibinfo
  {volume} {advance online publication}},\ \href
  {https://doi.org/10.1038/nphys3779} {10.1038/nphys3779}\BibitemShut {NoStop}%
\bibitem [{\citenamefont {Kim}\ \emph {et~al.}(2018)\citenamefont {Kim},
  \citenamefont {Schoenwaelder},\ and\ \citenamefont
  {Glenzer}}]{Kim2018development}%
  \BibitemOpen
  \bibfield  {author} {\bibinfo {author} {\bibfnamefont {J.~B.}\ \bibnamefont
  {Kim}}, \bibinfo {author} {\bibfnamefont {C.}~\bibnamefont {Schoenwaelder}},
  and\ \bibinfo {author} {\bibfnamefont {S.~H.}\ \bibnamefont {Glenzer}},\
  }\bibfield  {title} {\bibinfo {title} {Development and characterization of
  liquid argon and methane microjets for high-rep-rate laser-plasma
  experiments},\ }\href {https://doi.org/10.1063/1.5038561} {\bibfield
  {journal} {\bibinfo  {journal} {Review of Scientific Instruments}\ }\textbf
  {\bibinfo {volume} {89}},\ \bibinfo {pages} {10K105} (\bibinfo {year}
  {2018})}\BibitemShut {NoStop}%
\bibitem [{\citenamefont {Thoss}\ \emph {et~al.}(2003)\citenamefont {Thoss},
  \citenamefont {Richardson}, \citenamefont {Korn}, \citenamefont {Faubel},
  \citenamefont {Stiel}, \citenamefont {Vogt},\ and\ \citenamefont
  {Elsaesser}}]{thoss2003}%
  \BibitemOpen
  \bibfield  {author} {\bibinfo {author} {\bibfnamefont {A.}~\bibnamefont
  {Thoss}}, \bibinfo {author} {\bibfnamefont {M.}~\bibnamefont {Richardson}},
  \bibinfo {author} {\bibfnamefont {G.}~\bibnamefont {Korn}}, \bibinfo {author}
  {\bibfnamefont {M.}~\bibnamefont {Faubel}}, \bibinfo {author} {\bibfnamefont
  {H.}~\bibnamefont {Stiel}}, \bibinfo {author} {\bibfnamefont
  {U.}~\bibnamefont {Vogt}}, and\ \bibinfo {author} {\bibfnamefont
  {T.}~\bibnamefont {Elsaesser}},\ }\bibfield  {title} {{\selectlanguage
  {english}\bibinfo {title} {Kilohertz sources of hard x rays and fast ions
  with femtosecond laser plasmas}},\ }\href
  {https://doi.org/10.1364/JOSAB.20.000224} {\bibfield  {journal} {\bibinfo
  {journal} {JOSA B}\ }\textbf {\bibinfo {volume} {20}},\ \bibinfo {pages}
  {224} (\bibinfo {year} {2003})}\BibitemShut {NoStop}%
\bibitem [{\citenamefont {Inoue}(2016)}]{Inoue2016novel}%
  \BibitemOpen
  \bibfield  {author} {\bibinfo {author} {\bibfnamefont {N.~T.}\ \bibnamefont
  {Inoue}},\ }\bibfield  {title} {\bibinfo {title} {A novel microfluidic system
  for the mass production of inertial fusion energy shells},\ }\href
  {https://doi.org/10.1088/1742-6596/713/1/012011} {\bibfield  {journal}
  {\bibinfo  {journal} {Journal of Physics: Conference Series}\ }\textbf
  {\bibinfo {volume} {713}},\ \bibinfo {pages} {012011} (\bibinfo {year}
  {2016})}\BibitemShut {NoStop}%
\bibitem [{\citenamefont {Poole}\ \emph {et~al.}()\citenamefont {Poole},
  \citenamefont {Andereck}, \citenamefont {Schumacher}, \citenamefont
  {Daskalova}, \citenamefont {Feister}, \citenamefont {George}, \citenamefont
  {Willis}, \citenamefont {Akli},\ and\ \citenamefont
  {Chowdhury}}]{poole_liquid_2014}%
  \BibitemOpen
  \bibfield  {author} {\bibinfo {author} {\bibfnamefont {P.~L.}\ \bibnamefont
  {Poole}}, \bibinfo {author} {\bibfnamefont {C.~D.}\ \bibnamefont {Andereck}},
  \bibinfo {author} {\bibfnamefont {D.~W.}\ \bibnamefont {Schumacher}},
  \bibinfo {author} {\bibfnamefont {R.~L.}\ \bibnamefont {Daskalova}}, \bibinfo
  {author} {\bibfnamefont {S.}~\bibnamefont {Feister}}, \bibinfo {author}
  {\bibfnamefont {K.~M.}\ \bibnamefont {George}}, \bibinfo {author}
  {\bibfnamefont {C.}~\bibnamefont {Willis}}, \bibinfo {author} {\bibfnamefont
  {K.~U.}\ \bibnamefont {Akli}}, and\ \bibinfo {author} {\bibfnamefont {E.~A.}\
  \bibnamefont {Chowdhury}},\ }\bibfield  {title} {\bibinfo {title} {Liquid
  crystal films as on-demand, variable thickness (50–5000 nm) targets for
  intense lasers},\ }\href {https://doi.org/10.1063/1.4885100} {\ \textbf
  {\bibinfo {volume} {21}},\ \bibinfo {pages} {063109}}\BibitemShut {NoStop}%
\bibitem [{\citenamefont {Poole}\ \emph
  {et~al.}(2016{\natexlab{b}})\citenamefont {Poole}, \citenamefont {Willis},
  \citenamefont {Cochran}, \citenamefont {Hanna}, \citenamefont {Andereck},\
  and\ \citenamefont {Schumacher}}]{poole_moderate_2016}%
  \BibitemOpen
  \bibfield  {author} {\bibinfo {author} {\bibfnamefont {P.~L.}\ \bibnamefont
  {Poole}}, \bibinfo {author} {\bibfnamefont {C.}~\bibnamefont {Willis}},
  \bibinfo {author} {\bibfnamefont {G.~E.}\ \bibnamefont {Cochran}}, \bibinfo
  {author} {\bibfnamefont {R.~T.}\ \bibnamefont {Hanna}}, \bibinfo {author}
  {\bibfnamefont {C.~D.}\ \bibnamefont {Andereck}}, and\ \bibinfo {author}
  {\bibfnamefont {D.~W.}\ \bibnamefont {Schumacher}},\ }\bibfield  {title}
  {\bibinfo {title} {Moderate repetition rate ultra-intense laser targets and
  optics using variable thickness liquid crystal films},\ }\href
  {https://doi.org/10.1063/1.4964841} {\bibfield  {journal} {\bibinfo
  {journal} {Applied Physics Letters}\ }\textbf {\bibinfo {volume} {109}},\
  \bibinfo {pages} {151109} (\bibinfo {year} {2016}{\natexlab{b}})}\BibitemShut
  {NoStop}%
\bibitem [{\citenamefont {Gauthier}\ \emph {et~al.}(2017)\citenamefont
  {Gauthier}, \citenamefont {Curry}, \citenamefont {Göde}, \citenamefont
  {Brack}, \citenamefont {Kim}, \citenamefont {MacDonald}, \citenamefont
  {Metzkes}, \citenamefont {Obst}, \citenamefont {Rehwald}, \citenamefont
  {Rödel}, \citenamefont {Schlenvoigt}, \citenamefont {Schumaker},
  \citenamefont {Schramm}, \citenamefont {Zeil},\ and\ \citenamefont
  {Glenzer}}]{Gauthier2017high}%
  \BibitemOpen
  \bibfield  {author} {\bibinfo {author} {\bibfnamefont {M.}~\bibnamefont
  {Gauthier}}, \bibinfo {author} {\bibfnamefont {C.~B.}\ \bibnamefont {Curry}},
  \bibinfo {author} {\bibfnamefont {S.}~\bibnamefont {Göde}}, \bibinfo
  {author} {\bibfnamefont {F.-E.}\ \bibnamefont {Brack}}, \bibinfo {author}
  {\bibfnamefont {J.~B.}\ \bibnamefont {Kim}}, \bibinfo {author} {\bibfnamefont
  {M.~J.}\ \bibnamefont {MacDonald}}, \bibinfo {author} {\bibfnamefont
  {J.}~\bibnamefont {Metzkes}}, \bibinfo {author} {\bibfnamefont
  {L.}~\bibnamefont {Obst}}, \bibinfo {author} {\bibfnamefont {M.}~\bibnamefont
  {Rehwald}}, \bibinfo {author} {\bibfnamefont {C.}~\bibnamefont {Rödel}},
  \bibinfo {author} {\bibfnamefont {H.-P.}\ \bibnamefont {Schlenvoigt}},
  \bibinfo {author} {\bibfnamefont {W.}~\bibnamefont {Schumaker}}, \bibinfo
  {author} {\bibfnamefont {U.}~\bibnamefont {Schramm}}, \bibinfo {author}
  {\bibfnamefont {K.}~\bibnamefont {Zeil}}, and\ \bibinfo {author}
  {\bibfnamefont {S.~H.}\ \bibnamefont {Glenzer}},\ }\bibfield  {title}
  {\bibinfo {title} {High repetition rate, multi-{MeV} proton source from
  cryogenic hydrogen jets},\ }\href {https://doi.org/10.1063/1.4990487}
  {\bibfield  {journal} {\bibinfo  {journal} {Applied Physics Letters}\
  }\textbf {\bibinfo {volume} {111}},\ \bibinfo {pages} {114102} (\bibinfo
  {year} {2017})}\BibitemShut {NoStop}%
\bibitem [{\citenamefont {Kraft}\ \emph {et~al.}(2018)\citenamefont {Kraft},
  \citenamefont {Obst}, \citenamefont {Metzkes-Ng}, \citenamefont
  {Schlenvoigt}, \citenamefont {Zeil}, \citenamefont {Michaux}, \citenamefont
  {Chatain}, \citenamefont {Perin}, \citenamefont {Chen}, \citenamefont
  {Fuchs}, \citenamefont {Gauthier}, \citenamefont {Cowan},\ and\ \citenamefont
  {Schramm}}]{Kraft2018first}%
  \BibitemOpen
  \bibfield  {author} {\bibinfo {author} {\bibfnamefont {S.~D.}\ \bibnamefont
  {Kraft}}, \bibinfo {author} {\bibfnamefont {L.}~\bibnamefont {Obst}},
  \bibinfo {author} {\bibfnamefont {J.}~\bibnamefont {Metzkes-Ng}}, \bibinfo
  {author} {\bibfnamefont {H.-P.}\ \bibnamefont {Schlenvoigt}}, \bibinfo
  {author} {\bibfnamefont {K.}~\bibnamefont {Zeil}}, \bibinfo {author}
  {\bibfnamefont {S.}~\bibnamefont {Michaux}}, \bibinfo {author} {\bibfnamefont
  {D.}~\bibnamefont {Chatain}}, \bibinfo {author} {\bibfnamefont {J.-P.}\
  \bibnamefont {Perin}}, \bibinfo {author} {\bibfnamefont {S.~N.}\ \bibnamefont
  {Chen}}, \bibinfo {author} {\bibfnamefont {J.}~\bibnamefont {Fuchs}},
  \bibinfo {author} {\bibfnamefont {M.}~\bibnamefont {Gauthier}}, \bibinfo
  {author} {\bibfnamefont {T.~E.}\ \bibnamefont {Cowan}}, and\ \bibinfo
  {author} {\bibfnamefont {U.}~\bibnamefont {Schramm}},\ }\bibfield  {title}
  {\bibinfo {title} {First demonstration of multi-{MeV} proton acceleration
  from a cryogenic hydrogen ribbon target},\ }\href
  {https://doi.org/10.1088/1361-6587/aaae38} {\bibfield  {journal} {\bibinfo
  {journal} {Plasma Physics and Controlled Fusion}\ }\textbf {\bibinfo {volume}
  {60}},\ \bibinfo {pages} {044010} (\bibinfo {year} {2018})}\BibitemShut
  {NoStop}%
\bibitem [{\citenamefont {Komeda}\ \emph {et~al.}()\citenamefont {Komeda},
  \citenamefont {Nishimura}, \citenamefont {Mori}, \citenamefont {Hanayama},
  \citenamefont {Ishii}, \citenamefont {Okihara}, \citenamefont {Fujita},
  \citenamefont {Kitagawa}, \citenamefont {Sekine}, \citenamefont {Sato},
  \citenamefont {Kurita}, \citenamefont {Kawashima}, \citenamefont {Watari},
  \citenamefont {Kan}, \citenamefont {Nakamura}, \citenamefont {Kondo},
  \citenamefont {Fujine}, \citenamefont {Azuma}, \citenamefont {Motohiro},
  \citenamefont {Hioki}, \citenamefont {Kakeno}, \citenamefont {Sunahara},
  \citenamefont {Sentoku},\ and\ \citenamefont {Miura}}]{komeda_target_2013}%
  \BibitemOpen
  \bibfield  {author} {\bibinfo {author} {\bibfnamefont {O.}~\bibnamefont
  {Komeda}}, \bibinfo {author} {\bibfnamefont {Y.}~\bibnamefont {Nishimura}},
  \bibinfo {author} {\bibfnamefont {Y.}~\bibnamefont {Mori}}, \bibinfo {author}
  {\bibfnamefont {R.}~\bibnamefont {Hanayama}}, \bibinfo {author}
  {\bibfnamefont {K.}~\bibnamefont {Ishii}}, \bibinfo {author} {\bibfnamefont
  {S.}~\bibnamefont {Okihara}}, \bibinfo {author} {\bibfnamefont
  {K.}~\bibnamefont {Fujita}}, \bibinfo {author} {\bibfnamefont
  {Y.}~\bibnamefont {Kitagawa}}, \bibinfo {author} {\bibfnamefont
  {T.}~\bibnamefont {Sekine}}, \bibinfo {author} {\bibfnamefont
  {N.}~\bibnamefont {Sato}}, \bibinfo {author} {\bibfnamefont {T.}~\bibnamefont
  {Kurita}}, \bibinfo {author} {\bibfnamefont {T.}~\bibnamefont {Kawashima}},
  \bibinfo {author} {\bibfnamefont {T.}~\bibnamefont {Watari}}, \bibinfo
  {author} {\bibfnamefont {H.}~\bibnamefont {Kan}}, \bibinfo {author}
  {\bibfnamefont {N.}~\bibnamefont {Nakamura}}, \bibinfo {author}
  {\bibfnamefont {T.}~\bibnamefont {Kondo}}, \bibinfo {author} {\bibfnamefont
  {M.}~\bibnamefont {Fujine}}, \bibinfo {author} {\bibfnamefont
  {H.}~\bibnamefont {Azuma}}, \bibinfo {author} {\bibfnamefont
  {T.}~\bibnamefont {Motohiro}}, \bibinfo {author} {\bibfnamefont
  {T.}~\bibnamefont {Hioki}}, \bibinfo {author} {\bibfnamefont
  {M.}~\bibnamefont {Kakeno}}, \bibinfo {author} {\bibfnamefont
  {A.}~\bibnamefont {Sunahara}}, \bibinfo {author} {\bibfnamefont
  {Y.}~\bibnamefont {Sentoku}}, and\ \bibinfo {author} {\bibfnamefont
  {E.}~\bibnamefont {Miura}},\ }\bibfield  {title} {\bibinfo {title} {Target
  injection and engagement for neutron generation at 1 hz},\ }\href
  {https://doi.org/10.1585/pfr.8.1205020} {\ \textbf {\bibinfo {volume} {8}},\
  \bibinfo {pages} {1205020}}\BibitemShut {NoStop}%
\bibitem [{\citenamefont {Noaman-ul Haq}\ \emph {et~al.}()\citenamefont
  {Noaman-ul Haq}, \citenamefont {Ahmed}, \citenamefont {Sokollik},
  \citenamefont {Yu}, \citenamefont {Liu}, \citenamefont {Yuan}, \citenamefont
  {Yuan}, \citenamefont {Mirzaie}, \citenamefont {Ge}, \citenamefont {Chen},\
  and\ \citenamefont {Zhang}}]{noaman-ul-haq_statistical_2017}%
  \BibitemOpen
  \bibfield  {author} {\bibinfo {author} {\bibfnamefont {M.}~\bibnamefont
  {Noaman-ul Haq}}, \bibinfo {author} {\bibfnamefont {H.}~\bibnamefont
  {Ahmed}}, \bibinfo {author} {\bibfnamefont {T.}~\bibnamefont {Sokollik}},
  \bibinfo {author} {\bibfnamefont {L.}~\bibnamefont {Yu}}, \bibinfo {author}
  {\bibfnamefont {Z.}~\bibnamefont {Liu}}, \bibinfo {author} {\bibfnamefont
  {X.}~\bibnamefont {Yuan}}, \bibinfo {author} {\bibfnamefont {F.}~\bibnamefont
  {Yuan}}, \bibinfo {author} {\bibfnamefont {M.}~\bibnamefont {Mirzaie}},
  \bibinfo {author} {\bibfnamefont {X.}~\bibnamefont {Ge}}, \bibinfo {author}
  {\bibfnamefont {L.}~\bibnamefont {Chen}}, and\ \bibinfo {author}
  {\bibfnamefont {J.}~\bibnamefont {Zhang}},\ }\bibfield  {title} {\bibinfo
  {title} {Statistical analysis of laser driven protons using a
  high-repetition-rate tape drive target system},\ }\href
  {https://doi.org/10.1103/PhysRevAccelBeams.20.041301} {\ \textbf {\bibinfo
  {volume} {20}},\ \bibinfo {pages} {041301}}\BibitemShut {NoStop}%
\bibitem [{\citenamefont {Hou}\ \emph {et~al.}()\citenamefont {Hou},
  \citenamefont {Nees}, \citenamefont {Easter}, \citenamefont {Davis},
  \citenamefont {Petrov}, \citenamefont {Thomas},\ and\ \citenamefont
  {Krushelnick}}]{hou_mev_2009}%
  \BibitemOpen
  \bibfield  {author} {\bibinfo {author} {\bibfnamefont {B.}~\bibnamefont
  {Hou}}, \bibinfo {author} {\bibfnamefont {J.}~\bibnamefont {Nees}}, \bibinfo
  {author} {\bibfnamefont {J.}~\bibnamefont {Easter}}, \bibinfo {author}
  {\bibfnamefont {J.}~\bibnamefont {Davis}}, \bibinfo {author} {\bibfnamefont
  {G.}~\bibnamefont {Petrov}}, \bibinfo {author} {\bibfnamefont
  {A.}~\bibnamefont {Thomas}}, and\ \bibinfo {author} {\bibfnamefont
  {K.}~\bibnamefont {Krushelnick}},\ }\bibfield  {title} {\bibinfo {title}
  {{MeV} proton beams generated by 3 {mJ} ultrafast laser pulses at 0.5
  {kHz}},\ }\href {https://doi.org/10.1063/1.3224180} {\ \textbf {\bibinfo
  {volume} {95}},\ \bibinfo {pages} {101503}}\BibitemShut {NoStop}%
\bibitem [{\citenamefont {Kimbrough}\ \emph {et~al.}(2010)\citenamefont
  {Kimbrough}, \citenamefont {Bell}, \citenamefont {Bradley}, \citenamefont
  {Holder}, \citenamefont {Kalantar}, \citenamefont {MacPhee},\ and\
  \citenamefont {Telford}}]{Kimbrough2010standard}%
  \BibitemOpen
  \bibfield  {author} {\bibinfo {author} {\bibfnamefont {J.~R.}\ \bibnamefont
  {Kimbrough}}, \bibinfo {author} {\bibfnamefont {P.~M.}\ \bibnamefont {Bell}},
  \bibinfo {author} {\bibfnamefont {D.~K.}\ \bibnamefont {Bradley}}, \bibinfo
  {author} {\bibfnamefont {J.~P.}\ \bibnamefont {Holder}}, \bibinfo {author}
  {\bibfnamefont {D.~K.}\ \bibnamefont {Kalantar}}, \bibinfo {author}
  {\bibfnamefont {A.~G.}\ \bibnamefont {MacPhee}}, and\ \bibinfo {author}
  {\bibfnamefont {S.}~\bibnamefont {Telford}},\ }\bibfield  {title} {\bibinfo
  {title} {Standard design for national ignition facility x-ray streak and
  framing cameras},\ }\href {https://doi.org/10.1063/1.3496990} {\bibfield
  {journal} {\bibinfo  {journal} {Review of Scientific Instruments}\ }\textbf
  {\bibinfo {volume} {81}},\ \bibinfo {pages} {10E530} (\bibinfo {year}
  {2010})}\BibitemShut {NoStop}%
\bibitem [{\citenamefont {Metzkes}\ \emph {et~al.}(2016)\citenamefont
  {Metzkes}, \citenamefont {Zeil}, \citenamefont {Kraft}, \citenamefont
  {Karsch}, \citenamefont {Sobiella}, \citenamefont {Rehwald}, \citenamefont
  {Obst}, \citenamefont {Schlenvoigt},\ and\ \citenamefont
  {Schramm}}]{metzkes_online_2016}%
  \BibitemOpen
  \bibfield  {author} {\bibinfo {author} {\bibfnamefont {J.}~\bibnamefont
  {Metzkes}}, \bibinfo {author} {\bibfnamefont {K.}~\bibnamefont {Zeil}},
  \bibinfo {author} {\bibfnamefont {S.~D.}\ \bibnamefont {Kraft}}, \bibinfo
  {author} {\bibfnamefont {L.}~\bibnamefont {Karsch}}, \bibinfo {author}
  {\bibfnamefont {M.}~\bibnamefont {Sobiella}}, \bibinfo {author}
  {\bibfnamefont {M.}~\bibnamefont {Rehwald}}, \bibinfo {author} {\bibfnamefont
  {L.}~\bibnamefont {Obst}}, \bibinfo {author} {\bibfnamefont {H.-P.}\
  \bibnamefont {Schlenvoigt}}, and\ \bibinfo {author} {\bibfnamefont
  {U.}~\bibnamefont {Schramm}},\ }\bibfield  {title} {\bibinfo {title} {An
  online, energy-resolving beam profile detector for laser-driven proton
  beams},\ }\href {https://doi.org/10.1063/1.4961576} {\bibfield  {journal}
  {\bibinfo  {journal} {Review of Scientific Instruments}\ }\textbf {\bibinfo
  {volume} {87}},\ \bibinfo {pages} {083310} (\bibinfo {year}
  {2016})}\BibitemShut {NoStop}%
\bibitem [{\citenamefont {Reinhardt}\ \emph {et~al.}()\citenamefont
  {Reinhardt}, \citenamefont {Draxinger}, \citenamefont {Schreiber},\ and\
  \citenamefont {Assmann}}]{reinhardt_pixel_2013}%
  \BibitemOpen
  \bibfield  {author} {\bibinfo {author} {\bibfnamefont {S.}~\bibnamefont
  {Reinhardt}}, \bibinfo {author} {\bibfnamefont {W.}~\bibnamefont
  {Draxinger}}, \bibinfo {author} {\bibfnamefont {J.}~\bibnamefont
  {Schreiber}}, and\ \bibinfo {author} {\bibfnamefont {W.}~\bibnamefont
  {Assmann}},\ }\bibfield  {title} {\bibinfo {title} {A pixel detector system
  for laser-accelerated ion detection},\ }\href
  {https://doi.org/10.1088/1748-0221/8/03/P03008} {\ \textbf {\bibinfo {volume}
  {8}},\ \bibinfo {pages} {P03008}}\BibitemShut {NoStop}%
\bibitem [{\citenamefont {Prokůpek}\ \emph {et~al.}(2014)\citenamefont
  {Prokůpek}, \citenamefont {Kaufman}, \citenamefont {Margarone},
  \citenamefont {Krůs}, \citenamefont {Velyhan}, \citenamefont {Krása},
  \citenamefont {Burris-Mog}, \citenamefont {Busold}, \citenamefont {Deppert},
  \citenamefont {Cowan},\ and\ \citenamefont
  {Korn}}]{prokupek_development_2014}%
  \BibitemOpen
  \bibfield  {author} {\bibinfo {author} {\bibfnamefont {J.}~\bibnamefont
  {Prokůpek}}, \bibinfo {author} {\bibfnamefont {J.}~\bibnamefont {Kaufman}},
  \bibinfo {author} {\bibfnamefont {D.}~\bibnamefont {Margarone}}, \bibinfo
  {author} {\bibfnamefont {M.}~\bibnamefont {Krůs}}, \bibinfo {author}
  {\bibfnamefont {A.}~\bibnamefont {Velyhan}}, \bibinfo {author} {\bibfnamefont
  {J.}~\bibnamefont {Krása}}, \bibinfo {author} {\bibfnamefont
  {T.}~\bibnamefont {Burris-Mog}}, \bibinfo {author} {\bibfnamefont
  {S.}~\bibnamefont {Busold}}, \bibinfo {author} {\bibfnamefont
  {O.}~\bibnamefont {Deppert}}, \bibinfo {author} {\bibfnamefont {T.~E.}\
  \bibnamefont {Cowan}}, and\ \bibinfo {author} {\bibfnamefont
  {G.}~\bibnamefont {Korn}},\ }\bibfield  {title} {\bibinfo {title}
  {Development and first experimental tests of {Faraday} cup array},\ }\href
  {https://doi.org/10.1063/1.4859496} {\bibfield  {journal} {\bibinfo
  {journal} {Review of Scientific Instruments}\ }\textbf {\bibinfo {volume}
  {85}},\ \bibinfo {pages} {013302} (\bibinfo {year} {2014})}\BibitemShut
  {NoStop}%
\bibitem [{\citenamefont {Morrison}\ \emph {et~al.}(2018)\citenamefont
  {Morrison}, \citenamefont {Feister}, \citenamefont {Frische}, \citenamefont
  {Austin}, \citenamefont {Ngirmang}, \citenamefont {Murphy}, \citenamefont
  {{Chris Orban}}, \citenamefont {Chowdhury},\ and\ \citenamefont
  {Roquemore}}]{morrison_mev_2018}%
  \BibitemOpen
  \bibfield  {author} {\bibinfo {author} {\bibfnamefont {J.~T.}\ \bibnamefont
  {Morrison}}, \bibinfo {author} {\bibfnamefont {S.}~\bibnamefont {Feister}},
  \bibinfo {author} {\bibfnamefont {K.~D.}\ \bibnamefont {Frische}}, \bibinfo
  {author} {\bibfnamefont {D.~R.}\ \bibnamefont {Austin}}, \bibinfo {author}
  {\bibfnamefont {G.~K.}\ \bibnamefont {Ngirmang}}, \bibinfo {author}
  {\bibfnamefont {N.~R.}\ \bibnamefont {Murphy}}, \bibinfo {author}
  {\bibnamefont {{Chris Orban}}}, \bibinfo {author} {\bibfnamefont {E.~A.}\
  \bibnamefont {Chowdhury}}, and\ \bibinfo {author} {\bibfnamefont {W.~M.}\
  \bibnamefont {Roquemore}},\ }\bibfield  {title} {{\selectlanguage
  {english}\bibinfo {title} {{MeV} proton acceleration at {kHz} repetition rate
  from ultra-intense laser liquid interaction}},\ }\href
  {https://doi.org/10.1088/1367-2630/aaa8d1} {\bibfield  {journal} {\bibinfo
  {journal} {New Journal of Physics}\ }\textbf {\bibinfo {volume} {20}},\
  \bibinfo {pages} {022001} (\bibinfo {year} {2018})}\BibitemShut {NoStop}%
\bibitem [{\citenamefont {Dover}\ \emph {et~al.}(2017)\citenamefont {Dover},
  \citenamefont {Nishiuchi}, \citenamefont {Sakaki}, \citenamefont {Alkhimova},
  \citenamefont {Faenov}, \citenamefont {Fukuda}, \citenamefont {Kiriyama},
  \citenamefont {Kon}, \citenamefont {Kondo}, \citenamefont {Nishitani},
  \citenamefont {Ogura}, \citenamefont {Pikuz}, \citenamefont {Pirozhkov},
  \citenamefont {Sagisaka}, \citenamefont {Kando},\ and\ \citenamefont
  {Kondo}}]{dover_scintillator_2017}%
  \BibitemOpen
  \bibfield  {author} {\bibinfo {author} {\bibfnamefont {N.~P.}\ \bibnamefont
  {Dover}}, \bibinfo {author} {\bibfnamefont {M.}~\bibnamefont {Nishiuchi}},
  \bibinfo {author} {\bibfnamefont {H.}~\bibnamefont {Sakaki}}, \bibinfo
  {author} {\bibfnamefont {M.~A.}\ \bibnamefont {Alkhimova}}, \bibinfo {author}
  {\bibfnamefont {A.~Y.}\ \bibnamefont {Faenov}}, \bibinfo {author}
  {\bibfnamefont {Y.}~\bibnamefont {Fukuda}}, \bibinfo {author} {\bibfnamefont
  {H.}~\bibnamefont {Kiriyama}}, \bibinfo {author} {\bibfnamefont
  {A.}~\bibnamefont {Kon}}, \bibinfo {author} {\bibfnamefont {K.}~\bibnamefont
  {Kondo}}, \bibinfo {author} {\bibfnamefont {K.}~\bibnamefont {Nishitani}},
  \bibinfo {author} {\bibfnamefont {K.}~\bibnamefont {Ogura}}, \bibinfo
  {author} {\bibfnamefont {T.~A.}\ \bibnamefont {Pikuz}}, \bibinfo {author}
  {\bibfnamefont {A.~S.}\ \bibnamefont {Pirozhkov}}, \bibinfo {author}
  {\bibfnamefont {A.}~\bibnamefont {Sagisaka}}, \bibinfo {author}
  {\bibfnamefont {M.}~\bibnamefont {Kando}}, and\ \bibinfo {author}
  {\bibfnamefont {K.}~\bibnamefont {Kondo}},\ }\bibfield  {title} {\bibinfo
  {title} {Scintillator-based transverse proton beam profiler for laser-plasma
  ion sources},\ }\href {https://doi.org/10.1063/1.4994732} {\bibfield
  {journal} {\bibinfo  {journal} {Review of Scientific Instruments}\ }\textbf
  {\bibinfo {volume} {88}},\ \bibinfo {pages} {073304} (\bibinfo {year}
  {2017})}\BibitemShut {NoStop}%
\bibitem [{\citenamefont {Cirrone}\ \emph {et~al.}(2015)\citenamefont
  {Cirrone}, \citenamefont {Romano}, \citenamefont {Scuderi}, \citenamefont
  {Amato}, \citenamefont {Candiano}, \citenamefont {Cuttone}, \citenamefont
  {Giove}, \citenamefont {Korn}, \citenamefont {Krasa}, \citenamefont {Leanza},
  \citenamefont {Manna}, \citenamefont {Maggiore}, \citenamefont {Marchese},
  \citenamefont {Margarone}, \citenamefont {Milluzzo}, \citenamefont
  {Petringa}, \citenamefont {Sabini}, \citenamefont {Schillaci}, \citenamefont
  {Tramontana}, \citenamefont {Valastro},\ and\ \citenamefont
  {Velyhan}}]{cirrone_transport_2015}%
  \BibitemOpen
  \bibfield  {author} {\bibinfo {author} {\bibfnamefont {G.~A.~P.}\
  \bibnamefont {Cirrone}}, \bibinfo {author} {\bibfnamefont {F.}~\bibnamefont
  {Romano}}, \bibinfo {author} {\bibfnamefont {V.}~\bibnamefont {Scuderi}},
  \bibinfo {author} {\bibfnamefont {A.}~\bibnamefont {Amato}}, \bibinfo
  {author} {\bibfnamefont {G.}~\bibnamefont {Candiano}}, \bibinfo {author}
  {\bibfnamefont {G.}~\bibnamefont {Cuttone}}, \bibinfo {author} {\bibfnamefont
  {D.}~\bibnamefont {Giove}}, \bibinfo {author} {\bibfnamefont
  {G.}~\bibnamefont {Korn}}, \bibinfo {author} {\bibfnamefont {J.}~\bibnamefont
  {Krasa}}, \bibinfo {author} {\bibfnamefont {R.}~\bibnamefont {Leanza}},
  \bibinfo {author} {\bibfnamefont {R.}~\bibnamefont {Manna}}, \bibinfo
  {author} {\bibfnamefont {M.}~\bibnamefont {Maggiore}}, \bibinfo {author}
  {\bibfnamefont {V.}~\bibnamefont {Marchese}}, \bibinfo {author}
  {\bibfnamefont {D.}~\bibnamefont {Margarone}}, \bibinfo {author}
  {\bibfnamefont {G.}~\bibnamefont {Milluzzo}}, \bibinfo {author}
  {\bibfnamefont {G.}~\bibnamefont {Petringa}}, \bibinfo {author}
  {\bibfnamefont {M.~G.}\ \bibnamefont {Sabini}}, \bibinfo {author}
  {\bibfnamefont {F.}~\bibnamefont {Schillaci}}, \bibinfo {author}
  {\bibfnamefont {A.}~\bibnamefont {Tramontana}}, \bibinfo {author}
  {\bibfnamefont {L.}~\bibnamefont {Valastro}}, and\ \bibinfo {author}
  {\bibfnamefont {A.}~\bibnamefont {Velyhan}},\ }\bibfield  {title} {\bibinfo
  {title} {Transport and dosimetric solutions for the {ELIMED} laser-driven
  beam line},\ }\href {https://doi.org/10.1016/j.nima.2015.02.019} {\bibfield
  {journal} {\bibinfo  {journal} {Nuclear Instruments and Methods in Physics
  Research Section A: Accelerators, Spectrometers, Detectors and Associated
  Equipment}\ }\bibinfo {series} {Proceedings of the 10th {International}
  {Conference} on {Radiation} {Effects} on {Semiconductor} {Materials}
  {Detectors} and {Devices}},\ \textbf {\bibinfo {volume} {796}},\ \bibinfo
  {pages} {99} (\bibinfo {year} {2015})}\BibitemShut {NoStop}%
\bibitem [{\citenamefont {Everson}\ \emph {et~al.}(2009)\citenamefont
  {Everson}, \citenamefont {Pribyl}, \citenamefont {Constantin}, \citenamefont
  {Zylstra}, \citenamefont {Schaeffer}, \citenamefont {Kugland},\ and\
  \citenamefont {Niemann}}]{Everson2009design}%
  \BibitemOpen
  \bibfield  {author} {\bibinfo {author} {\bibfnamefont {E.~T.}\ \bibnamefont
  {Everson}}, \bibinfo {author} {\bibfnamefont {P.}~\bibnamefont {Pribyl}},
  \bibinfo {author} {\bibfnamefont {C.~G.}\ \bibnamefont {Constantin}},
  \bibinfo {author} {\bibfnamefont {A.}~\bibnamefont {Zylstra}}, \bibinfo
  {author} {\bibfnamefont {D.}~\bibnamefont {Schaeffer}}, \bibinfo {author}
  {\bibfnamefont {N.~L.}\ \bibnamefont {Kugland}}, and\ \bibinfo {author}
  {\bibfnamefont {C.}~\bibnamefont {Niemann}},\ }\bibfield  {title} {\bibinfo
  {title} {Design, construction, and calibration of a three-axis,
  high-frequency magnetic probe „b-dot probe{\ldots} as a diagnostic for
  exploding plasmas},\ }\href@noop {} {\bibfield  {journal} {\bibinfo
  {journal} {Rev. Sci. Insturm.}\ }\textbf {\bibinfo {volume} {80}},\ \bibinfo
  {pages} {113505} (\bibinfo {year} {2009})}\BibitemShut {NoStop}%
\bibitem [{\citenamefont {Heuer}\ \emph {et~al.}(2017)\citenamefont {Heuer},
  \citenamefont {Schaeffer}, \citenamefont {Knall}, \citenamefont {Constantin},
  \citenamefont {Hofer}, \citenamefont {Vincena}, \citenamefont {Tripathi},\
  and\ \citenamefont {Niemann}}]{Heuer2016fast}%
  \BibitemOpen
  \bibfield  {author} {\bibinfo {author} {\bibfnamefont {P.}~\bibnamefont
  {Heuer}}, \bibinfo {author} {\bibfnamefont {D.}~\bibnamefont {Schaeffer}},
  \bibinfo {author} {\bibfnamefont {E.}~\bibnamefont {Knall}}, \bibinfo
  {author} {\bibfnamefont {C.}~\bibnamefont {Constantin}}, \bibinfo {author}
  {\bibfnamefont {L.}~\bibnamefont {Hofer}}, \bibinfo {author} {\bibfnamefont
  {S.}~\bibnamefont {Vincena}}, \bibinfo {author} {\bibfnamefont
  {S.}~\bibnamefont {Tripathi}}, and\ \bibinfo {author} {\bibfnamefont
  {C.}~\bibnamefont {Niemann}},\ }\bibfield  {title} {\bibinfo {title} {Fast
  gated imaging of the collisionless interaction of a laser-produced and
  magnetized ambient plasma},\ }\href
  {https://doi.org/http://dx.doi.org/10.1016/j.hedp.2016.12.003} {\bibfield
  {journal} {\bibinfo  {journal} {High Energy Density Physics}\ }\textbf
  {\bibinfo {volume} {22}},\ \bibinfo {pages} {17 } (\bibinfo {year}
  {2017})}\BibitemShut {NoStop}%
\bibitem [{\citenamefont {Wilk}\ \emph {et~al.}()\citenamefont {Wilk},
  \citenamefont {Hochrein}, \citenamefont {Koch}, \citenamefont {Mei},\ and\
  \citenamefont {Holzwarth}}]{wilk_oscat_2011}%
  \BibitemOpen
  \bibfield  {author} {\bibinfo {author} {\bibfnamefont {R.}~\bibnamefont
  {Wilk}}, \bibinfo {author} {\bibfnamefont {T.}~\bibnamefont {Hochrein}},
  \bibinfo {author} {\bibfnamefont {M.}~\bibnamefont {Koch}}, \bibinfo {author}
  {\bibfnamefont {M.}~\bibnamefont {Mei}}, and\ \bibinfo {author}
  {\bibfnamefont {R.}~\bibnamefont {Holzwarth}},\ }\bibfield  {title} {\bibinfo
  {title} {{OSCAT}: Novel technique for time-resolved experiments without
  moveable optical delay lines},\ }\href
  {https://doi.org/10.1007/s10762-010-9670-8} {\ \textbf {\bibinfo {volume}
  {32}},\ \bibinfo {pages} {596}}\BibitemShut {NoStop}%
\bibitem [{\citenamefont {Hickstein}\ \emph {et~al.}()\citenamefont
  {Hickstein}, \citenamefont {Dollar}, \citenamefont {Ellis}, \citenamefont
  {Schnitzenbaumer}, \citenamefont {Keister}, \citenamefont {Petrov},
  \citenamefont {Ding}, \citenamefont {Palm}, \citenamefont {Gaffney},
  \citenamefont {Foord}, \citenamefont {Libby}, \citenamefont {Dukovic},
  \citenamefont {Jimenez}, \citenamefont {Kapteyn}, \citenamefont {Murnane},\
  and\ \citenamefont {Xiong}}]{hickstein_mapping_2014}%
  \BibitemOpen
  \bibfield  {author} {\bibinfo {author} {\bibfnamefont {D.~D.}\ \bibnamefont
  {Hickstein}}, \bibinfo {author} {\bibfnamefont {F.}~\bibnamefont {Dollar}},
  \bibinfo {author} {\bibfnamefont {J.~L.}\ \bibnamefont {Ellis}}, \bibinfo
  {author} {\bibfnamefont {K.~J.}\ \bibnamefont {Schnitzenbaumer}}, \bibinfo
  {author} {\bibfnamefont {K.~E.}\ \bibnamefont {Keister}}, \bibinfo {author}
  {\bibfnamefont {G.~M.}\ \bibnamefont {Petrov}}, \bibinfo {author}
  {\bibfnamefont {C.}~\bibnamefont {Ding}}, \bibinfo {author} {\bibfnamefont
  {B.~B.}\ \bibnamefont {Palm}}, \bibinfo {author} {\bibfnamefont {J.~A.}\
  \bibnamefont {Gaffney}}, \bibinfo {author} {\bibfnamefont {M.~E.}\
  \bibnamefont {Foord}}, \bibinfo {author} {\bibfnamefont {S.~B.}\ \bibnamefont
  {Libby}}, \bibinfo {author} {\bibfnamefont {G.}~\bibnamefont {Dukovic}},
  \bibinfo {author} {\bibfnamefont {J.~L.}\ \bibnamefont {Jimenez}}, \bibinfo
  {author} {\bibfnamefont {H.~C.}\ \bibnamefont {Kapteyn}}, \bibinfo {author}
  {\bibfnamefont {M.~M.}\ \bibnamefont {Murnane}}, and\ \bibinfo {author}
  {\bibfnamefont {W.}~\bibnamefont {Xiong}},\ }\bibfield  {title} {\bibinfo
  {title} {Mapping nanoscale absorption of femtosecond laser pulses using
  plasma explosion imaging},\ }\href {https://doi.org/10.1021/nn503199v} {\
  \textbf {\bibinfo {volume} {8}},\ \bibinfo {pages} {8810}}\BibitemShut
  {NoStop}%
\bibitem [{\citenamefont {He}\ \emph {et~al.}()\citenamefont {He},
  \citenamefont {Hou}, \citenamefont {Gao}, \citenamefont {Lebailly},
  \citenamefont {Nees}, \citenamefont {Clarke}, \citenamefont {Krushelnick},\
  and\ \citenamefont {Thomas}}]{he_coherent_2015}%
  \BibitemOpen
  \bibfield  {author} {\bibinfo {author} {\bibfnamefont {Z.-H.}\ \bibnamefont
  {He}}, \bibinfo {author} {\bibfnamefont {B.}~\bibnamefont {Hou}}, \bibinfo
  {author} {\bibfnamefont {G.}~\bibnamefont {Gao}}, \bibinfo {author}
  {\bibfnamefont {V.}~\bibnamefont {Lebailly}}, \bibinfo {author}
  {\bibfnamefont {J.~A.}\ \bibnamefont {Nees}}, \bibinfo {author}
  {\bibfnamefont {R.}~\bibnamefont {Clarke}}, \bibinfo {author} {\bibfnamefont
  {K.}~\bibnamefont {Krushelnick}}, and\ \bibinfo {author} {\bibfnamefont
  {A.~G.~R.}\ \bibnamefont {Thomas}},\ }\bibfield  {title} {\bibinfo {title}
  {Coherent control of plasma dynamics by feedback-optimized wavefront
  manipulationa)},\ }\href {https://doi.org/10.1063/1.4921159} {\ \textbf
  {\bibinfo {volume} {22}},\ \bibinfo {pages} {056704}}\BibitemShut {NoStop}%
\bibitem [{\citenamefont {Hutton}\ \emph {et~al.}(2012)\citenamefont {Hutton},
  \citenamefont {Azevedo}, \citenamefont {Beeler}, \citenamefont
  {Bettenhausen}, \citenamefont {Bond}, \citenamefont {Casey}, \citenamefont
  {Liebman}, \citenamefont {Marsh}, \citenamefont {Pannell},\ and\
  \citenamefont {Warrick}}]{Hutton2012experiment}%
  \BibitemOpen
  \bibfield  {author} {\bibinfo {author} {\bibfnamefont {M.~S.}\ \bibnamefont
  {Hutton}}, \bibinfo {author} {\bibfnamefont {S.}~\bibnamefont {Azevedo}},
  \bibinfo {author} {\bibfnamefont {R.}~\bibnamefont {Beeler}}, \bibinfo
  {author} {\bibfnamefont {R.}~\bibnamefont {Bettenhausen}}, \bibinfo {author}
  {\bibfnamefont {E.}~\bibnamefont {Bond}}, \bibinfo {author} {\bibfnamefont
  {A.}~\bibnamefont {Casey}}, \bibinfo {author} {\bibfnamefont
  {J.}~\bibnamefont {Liebman}}, \bibinfo {author} {\bibfnamefont
  {A.}~\bibnamefont {Marsh}}, \bibinfo {author} {\bibfnamefont
  {T.}~\bibnamefont {Pannell}}, and\ \bibinfo {author} {\bibfnamefont
  {A.}~\bibnamefont {Warrick}},\ }\bibfield  {title} {\bibinfo {title}
  {Experiment archive, analysis, and visualization at the national ignition
  facility},\ }\href {https://doi.org/10.1016/j.fusengdes.2012.07.009}
  {\bibfield  {journal} {\bibinfo  {journal} {Fusion Engineering and Design}\
  }\textbf {\bibinfo {volume} {87}},\ \bibinfo {pages} {2087} (\bibinfo {year}
  {2012})}\BibitemShut {NoStop}%
\bibitem [{\citenamefont {Bird}(2011)}]{Bird2011computing}%
  \BibitemOpen
  \bibfield  {author} {\bibinfo {author} {\bibfnamefont {I.}~\bibnamefont
  {Bird}},\ }\bibfield  {title} {\bibinfo {title} {Computing for the large
  hadron collider},\ }\href
  {https://doi.org/10.1146/annurev-nucl-102010-130059} {\bibfield  {journal}
  {\bibinfo  {journal} {Annual Review of Nuclear and Particle Science}\
  }\textbf {\bibinfo {volume} {61}},\ \bibinfo {pages} {99} (\bibinfo {year}
  {2011})}\BibitemShut {NoStop}%
\bibitem [{\citenamefont {Demchenko}\ \emph {et~al.}(2012)\citenamefont
  {Demchenko}, \citenamefont {Zhao}, \citenamefont {Grosso}, \citenamefont
  {Wibisono},\ and\ \citenamefont {de~Laat}}]{Demchenko2012addressing}%
  \BibitemOpen
  \bibfield  {author} {\bibinfo {author} {\bibfnamefont {Y.}~\bibnamefont
  {Demchenko}}, \bibinfo {author} {\bibfnamefont {Z.}~\bibnamefont {Zhao}},
  \bibinfo {author} {\bibfnamefont {P.}~\bibnamefont {Grosso}}, \bibinfo
  {author} {\bibfnamefont {A.}~\bibnamefont {Wibisono}}, and\ \bibinfo {author}
  {\bibfnamefont {C.}~\bibnamefont {de~Laat}},\ }\bibfield  {title} {\bibinfo
  {title} {Addressing big data challenges for scientific data infrastructure},\
  }in\ \href {https://doi.org/10.1109/cloudcom.2012.6427494} {\emph {\bibinfo
  {booktitle} {4th {IEEE} International Conference on Cloud Computing
  Technology and Science Proceedings}}}\ (\bibinfo  {publisher} {{IEEE}},\
  \bibinfo {year} {2012})\BibitemShut {NoStop}%
\bibitem [{\citenamefont {Katal}\ \emph {et~al.}(2013)\citenamefont {Katal},
  \citenamefont {Wazid},\ and\ \citenamefont {Goudar}}]{Katal2013big}%
  \BibitemOpen
  \bibfield  {author} {\bibinfo {author} {\bibfnamefont {A.}~\bibnamefont
  {Katal}}, \bibinfo {author} {\bibfnamefont {M.}~\bibnamefont {Wazid}}, and\
  \bibinfo {author} {\bibfnamefont {R.~H.}\ \bibnamefont {Goudar}},\ }\bibfield
   {title} {\bibinfo {title} {Big data: Issues, challenges, tools and good
  practices},\ }in\ \href {https://doi.org/10.1109/ic3.2013.6612229} {\emph
  {\bibinfo {booktitle} {2013 Sixth International Conference on Contemporary
  Computing ({IC}3)}}}\ (\bibinfo  {publisher} {{IEEE}},\ \bibinfo {year}
  {2013})\BibitemShut {NoStop}%
\bibitem [{\citenamefont {Group}(2010)}]{hdf5}%
  \BibitemOpen
  \bibfield  {author} {\bibinfo {author} {\bibfnamefont {T.~H.}\ \bibnamefont
  {Group}},\ }\href {http://www.hdfgroup.org/HDF5} {\bibinfo {title}
  {Hierarchical data format version 5}} (\bibinfo {year}
  {2000-2010})\BibitemShut {NoStop}%
\bibitem [{\citenamefont {Gizzi}\ \emph {et~al.}(2010)\citenamefont {Gizzi},
  \citenamefont {Clark}, \citenamefont {Neely}, \citenamefont {Roso},
  \citenamefont {Tolley}, \citenamefont {Gamucci}, \citenamefont {Giulietti},\
  and\ \citenamefont {Labate}}]{Gizzi2010high}%
  \BibitemOpen
  \bibfield  {author} {\bibinfo {author} {\bibfnamefont {L.~A.}\ \bibnamefont
  {Gizzi}}, \bibinfo {author} {\bibfnamefont {E.}~\bibnamefont {Clark}},
  \bibinfo {author} {\bibfnamefont {D.}~\bibnamefont {Neely}}, \bibinfo
  {author} {\bibfnamefont {L.}~\bibnamefont {Roso}}, \bibinfo {author}
  {\bibfnamefont {M.}~\bibnamefont {Tolley}}, \bibinfo {author} {\bibfnamefont
  {A.}~\bibnamefont {Gamucci}}, \bibinfo {author} {\bibfnamefont
  {A.}~\bibnamefont {Giulietti}}, and\ \bibinfo {author} {\bibfnamefont
  {L.}~\bibnamefont {Labate}},\ }\bibfield  {title} {\bibinfo {title} {High
  repetition rate laser systems: targets, diagnostics and radiation
  protection}\ }(\bibinfo  {publisher} {{AIP}},\ \bibinfo {year}
  {2010})\BibitemShut {NoStop}%
\end{thebibliography}%

\end{document}